\documentclass[aps,a4paper]{revtex4}
\usepackage{amssymb} \usepackage{amsmath,amssymb,graphicx,color,epsfig}
\usepackage{pstricks,float}
\usepackage[toc,page]{appendix}
\usepackage{tensor}
\usepackage{graphicx}   
\usepackage{caption}    
\usepackage{subcaption} 
\usepackage{subcaption}
\usepackage{subcaption}
\begin{document}
\title{Secondary Gravitational Wave Signatures from 5D Rotating Primordial Black Holes in the Dark Dimension.}
\author{Waqas Ahmed$^{1}$\footnote{E-mail: \texttt{\href{mailto:waqasmit@hbpu.edu.cn}{waqasmit@hbpu.edu.cn}}},
George K. Leontaris$^{2}$\footnote{E-mail: \texttt{\href{mailto:mjunaid@ualberta.ca}{leonta@uoi.gr}}} 
}

\affiliation{$^1$ Center for Fundamental Physics, School of Artificial Intelligence, Hubei Polytechnic University, Huangshi 435003, China.\\
$^2$ Physics Department, University of Ioannina, 45110, Ioannina, Greece.}


\begin{abstract}
We investigate five-dimensional rotating primordial black holes (PBHs) as dark matter candidates within the Dark Dimension (DD) scenario motivated by the Swampland Program. In this framework, a micron-scale extra dimension suppresses Hawking evaporation, allowing PBHs with initial masses \(M \gtrsim 10^{10}\,\mathrm{g}\) to survive to the present epoch. Moreover, the memory burden effect, a quantum-gravitational suppression of the evaporation rate by \(S^{-p}\), significantly prolongs PBH lifetimes and enlarges the allowed parameter space. We compute the evaporation dynamics for rotating 5D PBHs, derive the enhanced lifetime for \(p=2\), and establish the dark matter window \(10^{10}\,\mathrm{g} \lesssim M \lesssim 10^{21}\,\mathrm{g}\). The curvature perturbations responsible for PBH formation also generate a stochastic gravitational wave background through second-order scalar-induced effects. Assuming a log-normal primordial power spectrum with \(\sigma=1\) and \(f_{\mathrm{PBH}}=1\), we calculate the present-day energy density \(\Omega_{\mathrm{GW}}h^2\) across the Dark Dimension window. The predicted signals peak at frequencies from nHz to Hz, within the sensitivity ranges of LISA and DECIGO/BBO, while remaining consistent with current CMB spectral distortion bounds. Fisher forecasts show that future observatories can constrain the PBH mass, dark matter fraction, spectral width, and memory burden exponent with percent-level precision. A detection of the predicted gravitational wave background would provide simultaneous evidence for a micron-sized extra dimension, PBH dark matter, and the memory burden effect, offering a decisive test of quantum gravity and extra-dimensional physics.

\end{abstract}

\maketitle

\section{Introduction}

A wide range of astrophysical and cosmological observations confirm the existence of  dark matter in the universe, yet its fundamental nature remains unknown.
Primordial black holes (PBHs), formed from the gravitational collapse of overdense regions in the early universe~\cite{Carr:1974nx, Hawking:1971ei, Chapline:1975ojl, Carr:2023tpt}, have long been considered a compelling dark matter candidate.
In the standard four-dimensional framework, PBHs with masses
\[
M \lesssim 5 \times 10^{14}\,\mathrm{g}~,
\]
would have completely evaporated by the present day due to Hawking radiation~\cite{Hawking:1971ei}, and thus cannot be dark matter candidates. 
For higher masses, stringent bounds from the extragalactic gamma-ray background~\cite{Horowitz:2016lib} and constraints from the cosmic microwave background~\cite{Ricotti:2007au} severely limit their abundance, while microlensing limits~\cite{MACHO:2000qbb, Niikura:2017zjd} and gravitational wave merger rate studies~\cite{Sasaki:2016jop} make it highly improbable for PBHs in any sub-range to account for 100\% of the dark matter in four dimensions.

  	Renewed interest in PBHs has been driven by the detection of black-hole mergers via gravitational waves~\cite{LIGOScientific:2025slb},  
	(some of which may have a primordial origin),  together with significant theoretical progress on their formation mechanisms during
	 inflation and cosmological phase transitions.
  Theoretical progress including the higher dimensional (Dark Dimension) scenario, the memory burden effect or breakdown of semiclassical approximations,   has indicated  new mass windows for PBHs to constitute all or a significant fraction of the dark matter, by accounting for effects such as modified Hawking evaporation.

	Recent developments in the Swampland Program have opened new avenues for addressing these limitations. The Swampland Distance Conjecture, originally proposed by Ooguri and Vafa~\cite{Ooguri:2018wrx}, asserts that traversing a trans-Planckian distance in scalar field space leads to the emergence of an infinite tower of states whose masses become exponentially light. This light tower can undermine the validity and stability of the low-energy effective field theory. The conjecture, together with Bousso’s covariant entropy bound~\cite{Bousso:1999xy}, was shown to support a refined de Sitter conjecture and to extend naturally from AdS to de Sitter (or quasi-de Sitter) backgrounds\cite{Ooguri:2018wrx,Lust:2019zwm,Montero:2022prj}. 
    Applying this framework to the observed smallness of the dark energy scale $  \Lambda  $ naturally leads to the Dark Dimension (DD) scenario, which predicts a single compact extra dimension of micron-scale radius $  R_C  $ correlated with $  \Lambda  $:
%
\begin{equation}
R_C \sim \lambda \, \Lambda^{-1/4}, \quad \lambda \in [10^{-1}, 10^{-4}] \,,
\end{equation}

Fifth-force experiments~\cite{Adelberger:2003zx, Kapner:2006si} constrain the geometry, allowing only a single micron-size extra dimension, implying a KK mass scale 
\[
m_{KK} \gtrsim 6.6\,\mathrm{meV}
\]
and a five-dimensional Planck scale
\[
M_* \sim 10^{10}\,\mathrm{GeV}.
\]
The DD scenario profoundly impacts PBHs physics by suppressing  Hawking radiation  in higher dimensions, thus allowing lighter PBHs to survive to the present epoch~\cite{Carr:2020gox,Friedlander:2022ttk}. 
\\

In this work, we investigate the phenomenology of five-dimensional rotating primordial black holes (PBHs) within the framework of the Dark Dimension scenario. We extend previous studies  of five-dimensional Schwarzschild~\cite{Anchordoqui:2024akj, Anchordoqui:2024dxu} and rotating ~\cite{Leontaris:2025piz} PBHs by considering their potential role as dark matter and by analyzing in detail their Hawking evaporation, incorporating the memory burden effect. The latter is a quantum gravitational phenomenon in which the storage of information in PBHs suppresses Hawking evaporation~\cite{Dvali:2024hsb}.

We perform a systematic comparison between four- and five-dimensional dynamics and derive, in detail, the distinct mass windows for PBHs that can survive until the present epoch. Furthermore, we provide a complete derivation of the evaporation dynamics of five-dimensional rotating black holes, including the impact of the memory burden effect, and explore their connection to scalar-induced gravitational waves (SIGWs).

Our analysis also includes detailed numerical computations of the resulting stochastic gravitational wave background over a wide range of frequencies, from nHz to mHz. Finally, we explore the full parameter space in which PBHs can account for all or a fraction of the dark matter, demonstrating that even subdominant contributions can generate detectable gravitational wave signals.

The paper is organized as follows. In Sec.~II, we outline the theoretical framework, including the Swampland conjectures and the Dark Dimension scenario. Sec.~III reviews higher-dimensional black holes, focusing on both static and rotating solutions. In Sec.~IV, we analyze the memory burden effect and its impact on primordial black hole lifetimes. In Sec.~V we discuss primordial black hole formation from curvature perturbations while in Sec.~VI we derive the resulting scalar-induced gravitational wave spectrum. In  Sec.~VII we present our numerical results, followed by observational constraints in Sec.~VIII, and finally, in Sec.~IX we summarize  our conclusions and discuss future directions.

\section{Swampland Constraints and Dark Dimension Framework}

\subsection{The Swampland Program}

The Swampland Program aims to identify the subset of low-energy effective field theories that can arise from a consistent theory of quantum gravity \cite{Vafa:2005ui,Ooguri:2006in}. Within this framework, not every effective theory is considered viable; instead, a set of conjectured criteria is imposed to distinguish the consistent ``landscape'' from the inconsistent ``swampland.'' The latter consists of theories that, despite appearing consistent at low energies, cannot be embedded into a UV-complete framework such as string theory\cite{Palti:2019pca,vanBeest:2021lhn, Agmon:2022thq}.

A central element of this program is the \textit{Distance Conjecture}\cite{Ooguri:2018wrx}, which states that a large excursion in scalar field space leads to the appearance of an infinite tower of light states. For a field displacement $\Delta \phi \gg 1$ (in Planck units), the mass scale of this tower behaves as
\begin{equation}
m \sim m_0 \, e^{-\alpha \Delta \phi}, 
\qquad \alpha = \mathcal{O}(1).
\end{equation}

Another important constraint is the \textit{Weak Gravity Conjecture} (WGC)\cite{Ooguri:2018wrx, Arkani-Hamed:2006emk}, which asserts that gravity must be the weakest force in any consistent quantum gravity theory. For a $U(1)$ gauge theory with coupling $g$, this implies the existence of a charged state satisfying
\begin{equation}
\frac{m}{g M_P} \lesssim \mathcal{O}(1).
\end{equation}

\subsection{The AdS Distance Conjecture and Dark Energy}

The Distance Conjecture can be extended to curved backgrounds, particularly Anti-de Sitter (AdS) spacetime. The \textit{AdS Distance Conjecture}\cite{Montero:2022prj} suggests that as the cosmological constant $\Lambda < 0$ approaches zero, a tower of states becomes light. For the KK spectrum, this leads to
\begin{equation}
m_{\mathrm{KK}} \sim |\Lambda|^{a}, 
\qquad a = \frac{1}{4}.
\end{equation}

Using the observed value of the cosmological constant, one obtains
\begin{equation}
m_{\mathrm{KK}} \sim (10^{-122} M_P^4)^{1/4} 
\sim 10^{-30.5} M_P \sim \text{eV}.
\end{equation}
This corresponds to an extra dimension with characteristic size
\begin{equation}
R_C \sim \frac{1}{m_{\mathrm{KK}}} \sim \mu\text{m}.
\end{equation}

\subsection{Experimental Constraints}

Consistency conditions from quantum field theory in curved spacetime further constrain the parameter space. In particular, the Higuchi bound\cite{Higuchi:1986py} for massive spin-2 fields in de Sitter space restricts the exponent $a$ to the range\cite{Leontaris:2025piz}
\begin{equation}
\frac{1}{4} \leq a \leq \frac{1}{2}.
\end{equation}
Fifth-force experiments and precision tests of Newton's inverse-square law~\cite{Floratos:1999bv,Kehagias:1999my} further constrain the geometry of possible extra dimensions by placing strong upper bounds on deviations from $1/r$ gravity at sub-millimeter scales. In particular, such measurements restrict the compactification scale to the micron regime, thereby limiting the size of extra dimensions and the mass scale of the corresponding KK modes. These constraints can be interpreted as bounding the effective parameter space of the theory, effectively saturating the lower limit $a = 1/4$ in phenomenological realizations~\cite{Montero:2022prj,Anchordoqui:2023laz}. In the presence of compact extra dimensions, Newton's gravitational potential receives Yukawa-type corrections and can be written as
\begin{equation}
V(r) = -\frac{G_4 M_1 M_2}{r}
\left(1 + \alpha e^{-r/\lambda}\right).
\end{equation}
Experimental tests of the inverse-square law down to distances of order $30\,\mu\text{m}$\cite{Adelberger:2003zx, Lee:2020zjt} impose strong limits on such deviations. These results imply that any extra dimension must lie below this scale and strongly favor the existence of at most a single extra dimension in the micron range.

\subsection{The Species Scale}

The presence of a large number of light species lowers the effective scale at which quantum gravity becomes strongly coupled. This scale, known as the species scale\cite{Dvali:2007hz}, is particularly relevant in higher-dimensional setups.

For a single extra dimension ($n=1$), the relation between the four-dimensional Planck scale $M_P$ and the five-dimensional fundamental scale $M_*$ is given by
\begin{equation}
M_P^2 = M_*^3 R_C,
\end{equation}
where $R_C \sim 1/m_{\mathrm{KK}}$ is the compactification radius. This leads to
\begin{equation}
M_* \sim (m_{\mathrm{KK}} M_P^2)^{1/3} \sim 10^{10}\,\text{GeV}.
\end{equation}
This scale characterizes the onset of higher-dimensional gravitational effects.

\subsection{Implications for Primordial Black Holes}

The existence of a compact extra spatial dimension at the micron scale can have significant implications for PBH physics. In such higher-dimensional frameworks, the process of Hawking radiation is modified due to the altered gravitational phase space and the emergence of additional KK modes, which typically leads to a deviation from the standard four-dimensional evaporation rate~\cite{Hawking:1975vcx, Emparan:2000rs, Kanti:2004nr}. As a consequence, the lifetime of PBHs can be substantially affected, allowing lighter black holes to survive for longer cosmological timescales compared to the standard four-dimensional case~\cite{Carr:2016drx, Carr:2020gox}.


The phenomenological structure of PBHs also depends crucially on whether they are confined to a four-dimensional brane or allowed to probe the higher-dimensional bulk. In brane-localized scenarios, the emission spectrum is effectively restricted to Standard Model degrees of freedom on the brane, leading to comparatively faster evaporation. In contrast, bulk-propagating configurations allow additional gravitational channels, which modify the temperature–mass relation and can slow down the evaporation process~\cite{Emparan:2000rs, Cavaglia:2002si, Kanti:2004nr}. These differences naturally translate into distinct mass windows for PBHs that can survive until the present epoch.

For brane-localized black holes, the viable mass range is typically
\begin{equation}
10^{15} \lesssim \frac{M_{\mathrm{BH}}}{\mathrm{g}} \lesssim 10^{21},
\end{equation}
whereas in higher-dimensional or bulk-dominated scenarios the modified evaporation dynamics can relax the lower bound to
\begin{equation}
10^{10} \lesssim \frac{M_{\mathrm{BH}}}{\mathrm{g}} \lesssim 10^{21}.
\end{equation}
These differences arise from the modified Hawking temperature and enhanced emission channels in higher-dimensional gravity, which directly affect the evaporation timescale~\cite{Kanti:2004nr, Cavaglia:2002si}.

This framework is particularly relevant in light of recent developments connecting extra-dimensional physics with cosmological observations and quantum gravity constraints, where compact dimensions at sub-millimeter scales naturally emerge in phenomenological models addressing hierarchy and dark sector physics~\cite{Anchordoqui:2022txe, Anchordoqui:2024jkn, Anchordoqui:2024akj}.

\begin{figure}[t]
    \centering
    \includegraphics[width=0.8\columnwidth]{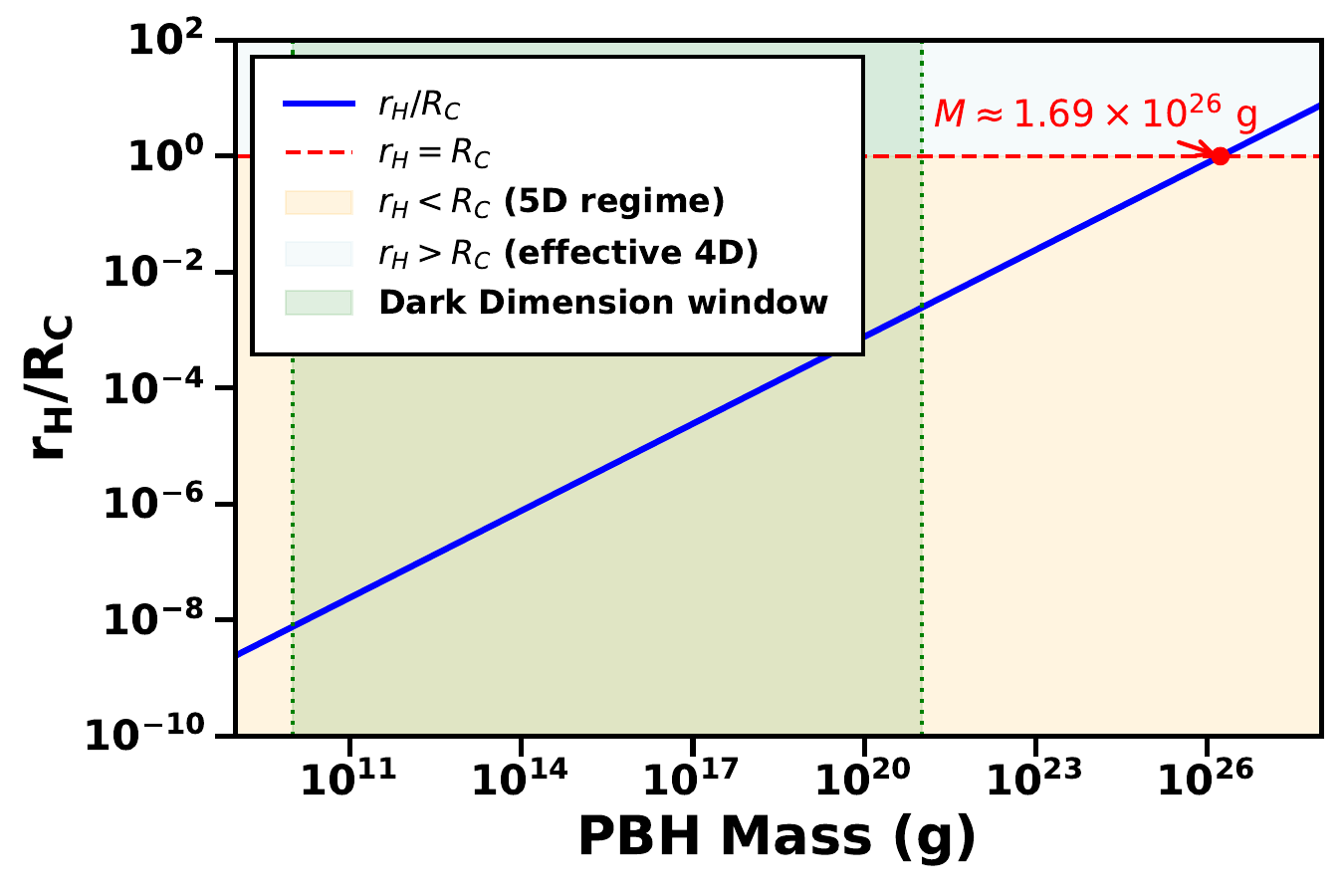}
    \caption{
        \label{fig:rH_over_RC}
     Ratio of the five‑dimensional Schwarzschild radius $r_H$ to the compactification radius $R_C$ as a function of the black hole mass $M$ (in grams).
        The 5D Planck scale is fixed to $M_* = 10^{10}\,\mathrm{GeV}$  and $R_C = 1\,\mu\mathrm{m}$.
        The red dashed line marks $r_H = R_C$.
        For $r_H \ll R_C$ (orange shaded region, left of the dashed line) the black hole is much smaller than the extra dimension; it does not “see” the compactness and behaves as a genuine five‑dimensional object (Tangherlini black hole). 
        For $r_H \gg R_C$ (light blue shaded region, right of the dashed line) the horizon wraps around the extra dimension, which is effectively averaged out, and the black hole follows the 4D Einstein equations (4D effective regime).
       The green shaded band highlights the Dark Dimension window 
$10^{10}\,\mathrm{g} \lesssim M \lesssim 10^{21}\,\mathrm{g}$ 
that is of particular interest for primordial black hole (PBH) dark matter candidates. 
The crossing point 
$M \approx 1.69 \times 10^{26}\,\mathrm{g}$ 
marks the transition between the two regimes.}
\end{figure}


\section{Higher-Dimensional Black Holes: Static and Rotating Solutions}

The spacetime geometry of a higher-dimensional black hole  critically depends on  the distance scale relative to the compactification radius \(R_C\). At distances much smaller than \(R_C\) (\(r \ll R_C\)), the gravitational field propagates in all \(4+n\) dimensions, and the black hole geometry is approximately that of a higher-dimensional Schwarzschild-Tangherlini solution \cite{Tangherlini:1963bw}:
\begin{equation}
ds^2 = -h(r) dt^2 + \frac{dr^2}{h(r)} + r^2 d\Omega_{n+2}^2,
\end{equation}
where
\begin{equation}
h(r) = 1 - \left(\frac{r_H}{r}\right)^{n+1},
\end{equation}
and \(r_H\) is the horizon radius, with \(d\Omega_{n+2}^2\) representing the metric on a unit \((n+2)\)-sphere. For one extra dimension (\(n=1\)), the horizon radius is \cite{Myers:1986un}:
\begin{equation}
r_S =\left(\frac{8}{3\pi} \,
\frac{M_{\rm BH}}{M_*}\right)^{\frac 12} \frac{1}{M_*} \label{rSofM}.
\end{equation}

 The ratio $r_H / R_C$ determines whether a black hole is effectively five‑dimensional ($r_H \ll R_C$) or appears four‑dimensional to brane‑localised observers ($r_H \gg R_C$).  
Figure~\ref{fig:rH_over_RC} displays this ratio as a function of the black hole mass $M$, using the 5D Planck scale \(M_* = 10^{10}\,\mathrm{GeV}\) and a compactification radius $R_C = 1\,\mu\mathrm{m}$ characteristic of the Dark Dimension scenario~\cite{Montero:2022prj}.  The shaded green band indicates the mass window \(10^{10}\,\mathrm{g} \lesssim M \lesssim 10^{21}\,\mathrm{g}\) that is phenomenologically relevant for primordial black hole dark matter candidates (see Section~\ref{sec:PBH_formation}).  
The crossing point $M \approx 1.69\times10^{26}\,\mathrm{g}$, where $r_H = R_C$, marks the boundary between the two geometric regimes.

\begin{figure}[t]
    \centering
    \includegraphics[width=0.8\columnwidth]{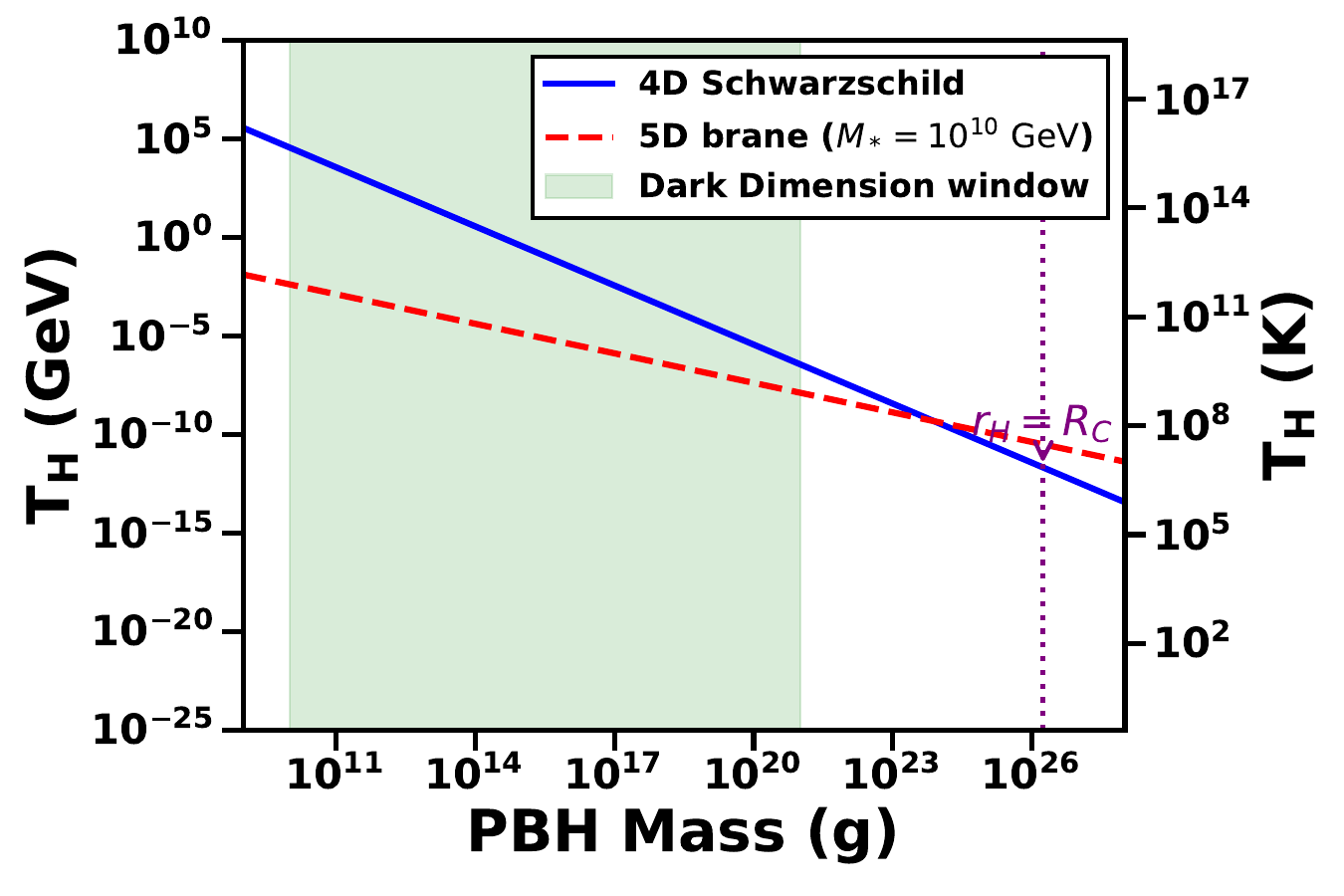}
    \caption{
        \label{fig:Hawking_T}
        Hawking temperature \(T_H\) as a function of the black hole mass \(M\).
The blue solid curve shows the standard 4D result $T_H \propto 1/M$, while the red dashed curve corresponds to a 5D brane‑localised black hole with $M_* = 10^{10}\,\mathrm{GeV}$.
The green shaded band marks the Dark Dimension window \(10^{10}\,\mathrm{g} \lesssim M \lesssim 10^{21}\,\mathrm{g}\).
The vertical purple line indicates the geometric transition mass \(M_{\text{cross}} \approx 1.69\times10^{26}\,\text{g}\) where \(r_H = R_C\). For masses much smaller than this, the black hole is in the 5D regime; for masses much larger, the extra dimension is averaged out and the black hole approaches 4D behaviour.
The temperature crossing (\(T_{\text{4D}} = T_{\text{5D}}\)) occurs at a much lower mass, \(M \sim 10^{20}\,\text{g}\), which lies within the right part of the Dark Dimension window.
For a reference mass of \(10^{15}\,\text{g}\), the 5D temperature is orders of magnitude lower than its 4D counterpart, illustrating the suppression of Hawking radiation in the 5D scenario.
The right vertical axis shows the temperature in Kelvin for convenience.
    }
\end{figure}


The Hawking temperature in \(4+n\) dimensions is
\begin{equation}
T_{BH} = \frac{n+1}{4\pi r_S},
\end{equation}
which for \(n=1\) reduces to
\begin{equation}
T_{BH} = \frac{1}{2\pi r_S}.
\end{equation}
Figure~\ref{fig:Hawking_T} compares the 4D and 5D temperatures. 
The 5D temperature is suppressed relative to the 4D case for masses above $\sim10^{21}\,\text{g}$ (the crossing point where $T_{\text{4D}} = T_{\text{5D}}$), while for lower masses the 5D temperature is actually higher.
The geometric transition at $r_H = R_C$ ($M \approx 1.69\times10^{26}\,\text{g}$) is indicated by the purple vertical line; note that this is far above the temperature crossing.
Both scales lie far above the Dark Dimension window $10^{10}\,\text{g} \lesssim M \lesssim 10^{21}\,\text{g}$, where the 5D temperature is strongly suppressed for most of the window (except the very low end). This suppression is the key feature that prolongs the lifetime of 5D primordial black holes and allows them to survive as dark matter candidates, especially when combined with the memory burden effect.

The entropy for \(n\) extra dimensions is
\begin{equation}
S = \frac{n+2}{n+1} M_{BH} T_{BH},
\end{equation}
and for \(n=1\):
\begin{equation}
S = \frac{3}{2} M_{BH} T_{BH}.
\end{equation}

For rotating black holes in five dimensions, the Myers-Perry solution \cite{Myers:1986un} provides the appropriate metric. In Boyer-Lindquist coordinates, assuming a single non-zero angular parameter aligned with the brane, the metric reduces to:
\begin{equation}
\begin{split}
ds^2 &= \left(1 - \frac{\mu}{\Sigma(r,\theta) r^{n-1}}\right) dt^2 + \frac{2 a \mu \sin^2 \theta}{\Sigma(r,\theta) r^{n-1}} dt d\phi - \frac{\Sigma(r,\theta)}{\Delta(r)} dr^2 \\
&\quad - \Sigma(r,\theta) d\theta^2 - \left(r^2 + a^2 + \frac{a^2 \mu \sin^2 \theta}{\Sigma(r,\theta) r^{n-1}}\right) \sin^2 \theta d\phi^2,
\end{split}
\end{equation}
where
\begin{equation}
\Delta(r) = r^2 + a^2 - \frac{\mu}{r^{n-1}}, \quad \Sigma(r,\theta) = r^2 + a^2 \cos^2 \theta.
\end{equation}
Here, \(\mu\) is related to the mass, and \(a = J/M\) is the rotation parameter. The event horizon satisfies
\begin{equation}
r_H^{\,n+1} (1 + a_*^2) = \mu, \quad a_* \equiv \frac{a}{r_H},
\end{equation}
so that
\begin{equation}
r_H = r_S (1 + a_*^2)^{1/(n+1)}, \quad r_S = \mu^{1/(n+1)}.
\end{equation}

The physical mass and angular momentum are
\begin{equation}
M_{BH} = \frac{(n+2) A_{n+2}}{16 \pi G_{4+n}} \mu, \quad J = \frac{2}{n+2} M_{BH} a,
\end{equation}
with \(A_{n+2} = \frac{2 \pi^{(n+3)/2}}{\Gamma\left(\frac{n+3}{2}\right)}\) being the area of a unit \((n+2)\)-sphere. For \(n=1\):
\begin{equation}
M_{BH} = \frac{3}{16 \pi G_5} A_3 \mu, \quad J = \frac{3}{2} M_{BH} a, \quad A_3 = 2 \pi^2.
\end{equation}

The Hawking temperature for a rotating black hole is
\begin{equation}
T_{BH} = \frac{1}{4 \pi r_H} \frac{n+1 + (n-1) a_*^2}{1 + a_*^2},
\end{equation}
and for \(n=1\):
\begin{equation}
T_{BH} = \frac{1}{2 \pi r_H (1 + a_*^2)}.
\end{equation}
The angular velocity at the horizon is
\begin{equation}
\Omega_H = \frac{a_*}{r_H (1 + a_*^2)}.
\end{equation}

The emission spectrum for a mode with quantum numbers \((s,\ell,m)\) is
\begin{equation}
\frac{d^3 E_{s \ell m}}{d\omega dt d\theta} = \frac{1}{2\pi} \frac{s \Gamma_{\ell m}(\omega) \, \omega}{\exp\left[(\omega - m \Omega)/T_H\right] - (-1)^{2s}} \int_0^{2\pi} |s S_{\ell m}(\theta, a \omega)|^2 d\phi,
\end{equation}
where the greybody factors \(\Gamma_{\ell m}^s(\omega)\) encode the probability of escape from the black hole potential, and the total mass loss rate is
\begin{equation}
-\frac{dM}{dt} = \frac{1}{2\pi} \sum_{s,\ell,m} g_s \int_0^\infty \frac{s \Gamma_{\ell m}(\omega) \, \omega}{\exp\left[(\omega - m \Omega)/T_H\right] - (-1)^{2s}} d\omega,
\end{equation}
with degrees of freedom \(g_0 = 4\) (scalars), \(g_{1/2} = 90\) (fermions), and \(g_1 = 24\) (vectors) at temperature \(T_H\). In the low-frequency regime \(\omega r_H \ll 1\), greybody factors can be computed analytically \cite{Kanti:2004nr}.

Finally, the dimensionless parameter
\begin{equation}
\tilde Q = \frac{\omega - m \Omega}{2 \pi T_H} = (1 + a_*^2) \tilde \omega - m a_*,
\end{equation}
controls the frequency dependence of the greybody factors. This unified framework sets the stage for discussing primordial black hole formation, evaporation, and evolution in higher-dimensional spacetimes.

\subsection{Evaporation Dynamics of Five-Dimensional Rotating Primordial Black Holes}

Primordial black holes with rotation present a richer evaporation dynamics compared to non-rotating counterparts. The evolution of both mass $M$ and angular momentum $J$ is coupled, with the time evolution governed by~\cite{Kanti:2002nr, Ida:2005ax}:
\begin{equation}
\frac{dM}{dt} = -C_M(a_*) r_H^2, \quad \frac{dJ}{dt} = -C_J(a_*) r_H,
\end{equation}
where $C_M(a_*)$ and $C_J(a_*)$ are coefficients derived from integrating greybody factors over all emission modes. Table~\ref{tab:rot_coeff} lists representative values for different dimensionless spin parameters $a_*$.

\begin{table}[h!]
\centering
\caption{Coefficients $C_M$ and $C_J$ for different rotation parameters $a_*$ and their ratio $\eta = C_M / C_J$.}
\label{tab:rot_coeff}
\begin{tabular}{c|c|c|c}
\hline
$a_*$ & $C_M$ & $C_J$ & $\eta$ \\
\hline
0.5 & 0.07 & 0.11 & 0.64 \\
1.0 & 0.11 & 0.22 & 0.50 \\
1.5 & 0.05 & 0.14 & 0.33 \\
\hline
\end{tabular}
\end{table}

Eliminating time between these equations yields a direct relation between mass and angular momentum:
\begin{equation}
\frac{dJ}{dM} = \frac{1}{\eta} r_H.
\end{equation}
For a five-dimensional rotating black hole ($n=1$) with $G_5 = 1/(8 \pi M_*^3)$, the horizon radius satisfies
\begin{equation}
r_H^2 = \kappa \left(1 - \lambda \frac{J^2}{\kappa}\right),
\end{equation}
where
\begin{equation}
\kappa = \frac{8 G_5 M^3}{\pi}, \quad \lambda = \frac{9}{4} M^2.
\end{equation}
The extremal angular momentum is therefore $J_{\rm max}^2 = \kappa / \lambda$, and a normalized spin parameter can be defined as $L = J / J_{\rm max}$ with $L \leq 1$. In terms of $L$, the horizon radius becomes
\begin{equation}
r_H = \pi M_* \sqrt{\frac{3 M}{M_*} (1 - L^2)}.
\end{equation}

\begin{figure}[htbp]
\centering
\includegraphics[width=0.7\textwidth]{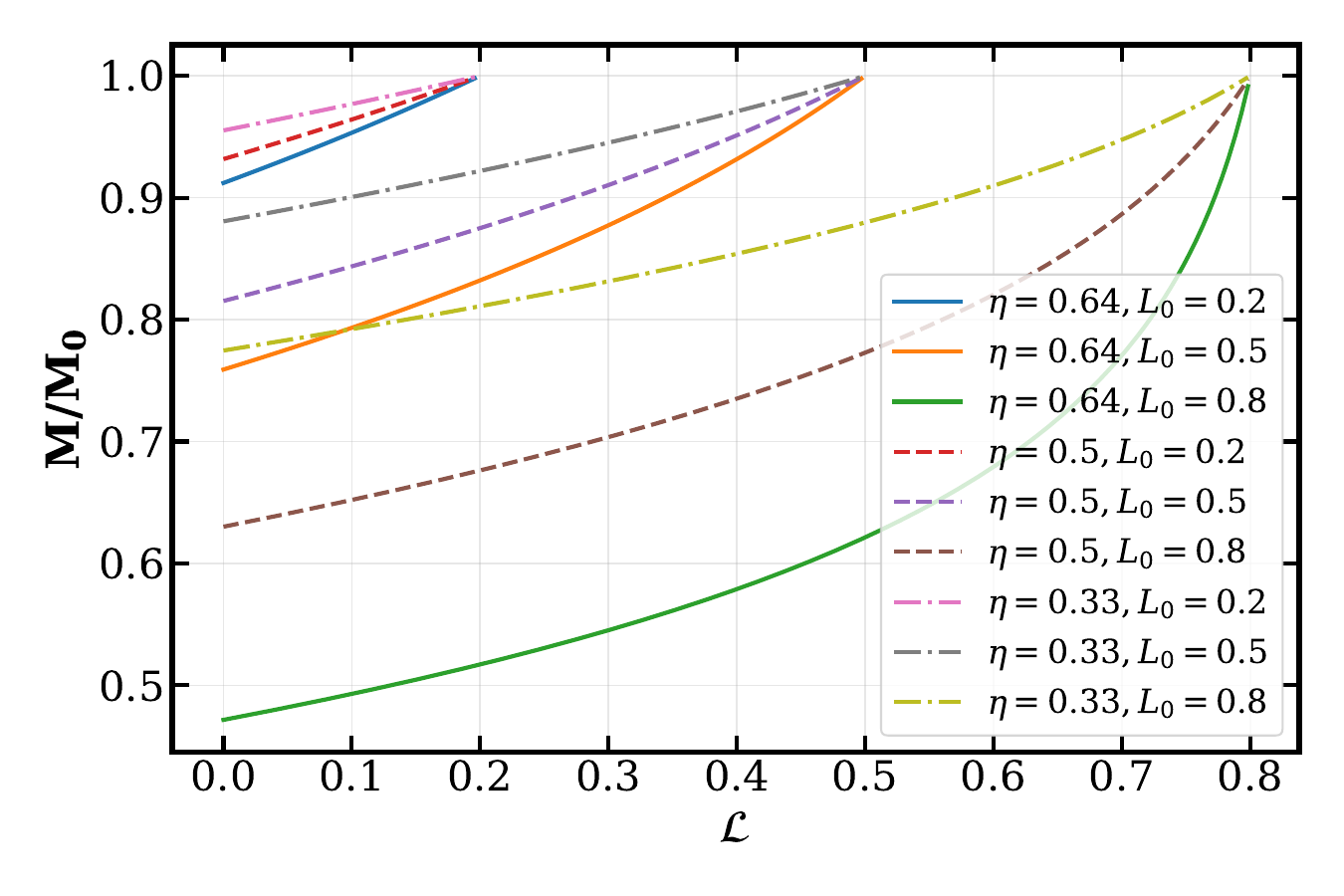}
\caption{Evolution of the mass fraction $M/M_0$ as a function of the normalized angular momentum $\mathcal{L}=J/J_{\text{max}}$ for five‑dimensional rotating black holes. Solid, dashed, and dash‑dotted lines correspond to $\eta = 0.64$, $0.50$, $0.33$ (i.e., rotation parameters $a_*=0.5$, $1.0$, $1.5$), while colours indicate different initial spins $\mathcal{L}_0$ ranging from $0.2$ to $0.95$. Higher initial spin results in a longer spin‑down phase and a larger mass loss (typically $40\%$–$60\%$ of the initial mass is radiated away before the black hole becomes non‑rotating).}
\label{fig:spindown}
\end{figure}

Substituting into the mass–angular momentum relation, the evolution equation for $L$ reads
\begin{equation}
\frac{dL}{1-L^2 - \eta L} = \frac{3}{2} \eta \frac{dM}{M}.
\end{equation}
Integrating gives an implicit relation between $M$ and $L$:
\begin{equation}\label{spinL}
f(L) - f(L_0) = \frac{2}{3} \left(\eta + \frac{1}{\eta}\right) \ln \frac{M}{M_0}, \quad
f(L) = \sin^{-1} L - \eta \ln (1 - L^2 - \eta L),
\end{equation}
which can be inverted to express $M$ as a function of $L$:
\begin{equation}
M(L) = M_0 \left( \frac{1 - L^2 - \eta L}{1 - L_0^2 - \eta L_0} \right)^{\eta \zeta} \exp\left[ \zeta (\sin^{-1} L - \sin^{-1} L_0) \right], \quad \zeta = \frac{3}{2} \eta.
\end{equation}
Figure~\ref{fig:spindown} illustrates the coupled evolution of mass and angular momentum during the spin‑down phase. Curves are obtained by numerically inverting the implicit relation for different $\eta$ and initial spins $\mathcal{L}_0$. Higher initial spin results in a longer spin‑down phase and larger mass loss (typically $40\%$–$60\%$ of the initial mass is radiated away). The spin‑down timescale is $\tau_{\text{sp}} \approx 5.2\times10^{-15} (M_i\,\text{g})^2$ years; for $M_i = 5\times10^{11}\,\text{g}$, $\tau_{\text{sp}} \approx 3.3\times10^9$ years.

Once the black hole reaches $L=0$, it enters the non‑rotating Schwarzschild phase. The remaining mass $M_f \sim 0.4-0.6 M_i$ has a lifetime $\tau_{\text{Sch}} \approx 9\times10^{-15} (M_f\,\text{g})^2$ years, giving $\tau_{\text{Sch}} \approx 2\times10^8$ years for $M_f = 1.5\times10^{11}\,\text{g}$. The total lifetime $\tau_{\text{total}} = \tau_{\text{sp}} + \tau_{\text{Sch}} \approx 3.5\times10^9$ years is comparable to the age of the Universe, demonstrating that 5D rotating PBHs with initial masses as low as $10^{11}\,\text{g}$ can survive to the present, unlike in 4D where the threshold is $M \gtrsim 5\times10^{14}\,\text{g}$.
These results highlight the crucial role of rotation in PBH evaporation dynamics, extending lifetimes and potentially affecting their contribution to the dark matter content of the Universe.
\subsection{Memory Burden and the Evaporation of Primordial Black Holes}

The standard semiclassical treatment of black hole evaporation, as derived by Hawking, predicts that black holes emit thermal radiation~\cite{Hawking:1975vcx}, seemingly erasing all information about the matter that formed them. This apparent loss of information leads to a direct conflict with the principles of quantum mechanics, a problem known as the black hole information paradox~\cite{Hawking:1976ra}. Addressing this paradox is crucial for understanding the ultimate fate of primordial black holes (PBHs) and their potential role as dark matter candidates~\cite{Carr:2016drx, Carr:2020gox}.

A compelling resolution has been proposed through the concept of the \emph{memory burden effect}~\cite{Dvali:2015aja,Dvali:2011aa}. In this framework, as a black hole loses mass through Hawking radiation, information about its initial state is not lost; instead, it becomes encoded in the remaining degrees of freedom of the black hole. This accumulated information effectively acts as a feedback mechanism, gradually suppressing the evaporation rate and stabilizing the black hole against complete dissipation~\cite{Dvali:2020wft}.

Mathematically, this behavior can be captured by a Hamiltonian describing a master mode and a set of memory modes~\cite{Dvali:2011aa}:
\begin{equation}
\hat{H} = \epsilon_0 \hat{n}_0 + \hat{E}_K \sum_{k=1}^{N_m} \hat{n}_k,
\end{equation}
where the energy gap of the memory modes depends on the occupation of the master mode~\cite{Dvali:2020wft}:
\begin{equation}
\hat{E}_K = \left(1 - \frac{\hat{n}_0}{N_c}\right)^p \epsilon_K.
\end{equation}
Here, $N_c$ represents the critical occupation number, related to the black hole's entropy $S$, $p$ is a parameter controlling the strength of the memory burden effect, and $\epsilon_0$ and $\epsilon_K$ are characteristic energy scales. At early stages of evaporation, when $\hat{n}_0 \ll N_c$, memory modes carry the full energy cost $\hat{E}_K \approx \epsilon_K$. As the black hole approaches the critical occupation number, $\hat{n}_0 \to N_c$, the energy cost vanishes, $\hat{E}_K \to 0$, creating an effective energy barrier that slows further evaporation.

Mapping these quantities onto black hole parameters yields
\begin{equation}
\epsilon_0 = r_S^{-1}, \quad N_c = S, \quad N_m = \frac{S}{2}, \quad \epsilon_K = \sqrt{S}\, r_S^{-1},
\end{equation}
with $r_S$ the Schwarzschild radius and $S$ the Bekenstein-Hawking entropy~\cite{Bekenstein:1973ur}. The onset of the memory burden phase occurs after a critical fraction of the black hole mass has been lost,
\begin{equation}\label{rate}
q \equiv \frac{\Delta M_{\rm crit}}{M_0} \sim (p S)^{-1/p},
\end{equation}
illustrating that the timing of this effect strongly depends on the parameter $p$. For $p=1$, the memory burden emerges almost immediately; for larger $p$, the onset is delayed until a significant portion of the black hole mass is radiated away.

The modified mass-loss equation incorporating the memory burden can be expressed as
\begin{equation}
\frac{dM}{dt} = -\frac{C_M}{S^p} r_H^2 M,
\end{equation}
which for a five-dimensional Schwarzschild black hole with appropriate substitutions becomes
\begin{equation}
\frac{dM}{dt} = -C_S M^{\frac{3p}{2}+1}, \quad
C_S = C_M \left(\frac{3}{4\pi}\right)^{-p} \left(\frac{3\pi}{8G_5}\right)^{-\frac{3p}{2}-1}.
\end{equation}
Integration of the modified evaporation equation reveals that the black hole lifetime is significantly extended relative to the standard Hawking prediction, with the lifetime depending sensitively on both the memory burden exponent $p$ and the initial mass. The general expression in natural units is
\begin{equation}\label{MMeff}
\tau^{(p)} = \frac{2^{2p+1}}{3^{\frac{3p}{2}+1}} \frac{1}{C_M \pi^2 (3p+4)} \left( \frac{M}{M_*} \right)^{\frac{3p}{2}+2} \frac{1}{M_*}.
\end{equation}
For the Dark Dimension scenario, the five‑dimensional Planck scale is $M_* = 10^{10}\,\mathrm{GeV}$ and the  coefficient is $C_M \approx 0.04$. Converting to years (using $\hbar$ and the number of seconds per year), we obtain the following numerical approximations:
\begin{align}
\tau^{(p=0)}(M) &= 2.78 \times 10^{-15} \left( \frac{M}{\mathrm{g}} \right)^2 \;\text{years}, \label{eq:tau_p0}\\
\tau^{(p=1)}(M) &= 5.15 \times 10^{7} \left( \frac{M}{\mathrm{g}} \right)^{7/2} \;\text{years}, \label{eq:tau_p1}\\
\tau^{(p=2)}(M) &= 1.17 \times 10^{26} \left( \frac{M}{\mathrm{g}} \right)^{5} \;\text{years}. \label{eq:tau_p2}
\end{align}
These formulas demonstrate that even relatively light black holes can remain stable over cosmological timescales. For example, a light PBH of mass $10^{10}\,\mathrm{g}$ can survive to the present day only for $p=2$, opening a previously excluded mass window for PBH dark matter. To illustrate, Table~\ref{tab:PBH_lifetime} compares lifetimes for PBHs of different masses under various evaporation scenarios, including the memory burden effect.
The detailed analysis of PBH formation is presented in the next section.

\begin{table}[h!]
\centering
\caption{PBH lifetime comparison under standard Hawking evaporation and with the memory burden (MB) effect in five dimensions. All lifetimes exceed the age of the Universe for $M\ge10^{10}\,\mathrm{g}$, but only the $p=2$ case avoids observational constraints from BBN and CMB spectral distortions.}
\label{tab:PBH_lifetime}
\begin{tabular}{c|c|c|c|c}
\hline
Mass (g) & 4D Hawking (yr) & 5D no MB (yr) & 5D MB $p=1$ (yr) & 5D MB $p=2$ (yr) \\
\hline
$10^{10}$ & $1.0 \times 10^{-5}$ & $2.8 \times 10^{5}$ & $5.2 \times 10^{42}$ & $1.2 \times 10^{76}$ \\
$10^{12}$ & $1.0 \times 10^{1}$  & $2.8 \times 10^{9}$ & $5.2 \times 10^{49}$ & $1.2 \times 10^{86}$ \\
$10^{15}$ & $1.0 \times 10^{10}$ & $2.8 \times 10^{15}$ & $5.2 \times 10^{59}$ & $1.2 \times 10^{101}$ \\
$10^{18}$ & $1.0 \times 10^{19}$ & $2.8 \times 10^{21}$ & $5.2 \times 10^{69}$ & $1.2 \times 10^{116}$ \\
\hline
\end{tabular}
\end{table}
\subsection{Energy Injection Constraints and the $p=1$ Exclusion}

A PBH can survive to the present day (a total lifetime $\tau_{\rm total} > t_U$) without necessarily being cosmologically viable. The crucial quantity is not the total lifetime, but the \textbf{rate of energy injection} $\dot{M}(t)$ in various cosmological epochs.

Even if a PBH survives until today, its Hawking emission during BBN ($t \sim 1$-$10^3$ s), or around recombination ($t \sim 10^{13}$ s) may still produce observable cosmological signatures, such as:
\begin{itemize}
\item alterations of the primordial light-element abundances (D, $^3$He, $^4$He, $^7$Li) through photodissociation and hadronic cascades; 
\item $\mu$-type and $y$-type spectral distortions of the CMB.
\end{itemize}

These constraints are well-established in the PBH literature~\cite{Carr:2020gox, Keith:2020jww, Korwar:2023kpy, Tashiro:2008sf, Acharya:2020jbv} and are independent of the specific PBH formation mechanism.

For $p=1$, the memory burden becomes active almost immediately after formation. From Eq.~\eqref{rate}, the critical mass fraction is $q \sim S^{-1} \ll 1$, which means that the memory-burdened phase begins when only a tiny fraction of the initial mass has evaporated. Although the evaporation rate in this phase is suppressed by $S^{-1}$ relative to the standard Hawking rate, the suppression is relatively modest. The lifetime scales as $\tau^{(p=1)} \propto M^{7/2}$ (Eq.~\ref{eq:tau_p1}), and the energy injection rate $\dot{M}(t)$ during BBN and recombination, while reduced, is still sufficiently large to fall within the observationally excluded regions. This is why the red dash-dotted curve in Fig.~\ref{fig:lifetime}, despite lying well above the age-of-Universe line, is excluded by the shaded BBN and CMB distortion constraints.

For $p=2$, the situation is qualitatively different. The onset of memory burden also occurs after an exponentially tiny mass loss, $q \sim (2S)^{-1/2} \sim 10^{-18}$. However, the suppression of the evaporation rate scales as $S^{-2}$ rather than $S^{-1}$, which is dramatically stronger. Consequently, even though the black hole enters the burdened phase with almost its full mass, its instantaneous luminosity is suppressed by a factor $\sim 10^{-70}$ relative to the standard 5D Hawking rate. The lifetime scales as $\tau^{(p=2)} \propto M^{5}$ (Eq.~\ref{eq:tau_p2}), and the energy injection rate during BBN and recombination is completely negligible, easily satisfying all observational bounds.

This is the fundamental physical reason why only the $p=2$ memory burden scenario opens a viable new dark matter window in the mass range $10^{10}\,\mathrm{g} \lesssim M \lesssim 10^{15}\,\mathrm{g}$. The extremely strong suppression ensures that energy injection constraints are satisfied while the integrated lifetime exceeds the age of the Universe.

\subsection{Primordial Black Hole Formation and Mass Distribution}\label{sec:PBH_formation}
Primordial black holes (PBHs) can form in the early Universe if sufficiently large curvature perturbations produced during inflation re-enter the Hubble horizon during the radiation-dominated epoch and undergo gravitational collapse~\cite{Zeldovich:1967lct,Hawking:1971ei,Carr:1974nx}. When a perturbation mode re-enters the horizon, the corresponding overdense region may collapse into a black hole if the density contrast exceeds a critical threshold. The density contrast is defined as
\begin{equation}
\delta = \frac{\delta \rho}{\rho},
\end{equation}
 whilst numerical simulations indicate that gravitational collapse occurs~\cite{Harada:2013epa,Kuhnel:2015vtw} when
\begin{equation}
\delta > \delta_c \simeq 0.45 
\end{equation}

The mass of the PBH at formation is expected to be comparable to the mass contained within the Hubble horizon at the time of re-entry. This can be expressed as
\begin{equation}
M_{\rm PBH} \simeq \gamma M_H
           = \gamma \frac{4\pi}{3}\rho H^{-3}\Big|_{\rm re-entry},
\end{equation}
where $\gamma \simeq 0.2$~\cite{Carr:2020gox,Carr:2009jm} characterizes the efficiency of the collapse process.

During the radiation-dominated epoch, the horizon mass is related to the Hubble parameter through~\cite{Carr:2020xqk}
\begin{equation}
M_H = \frac{1}{2G\,H}~\;; \qquad k = aH~.
\end{equation}
In the above, $k$ denotes the co-moving wavenumber of the perturbation mode. By combining entropy conservation with the temperature evolution of the Universe, we obtain a relation between the PBH mass and the corresponding comoving scale,
\begin{equation}\label{kstar}
k \simeq 7 \times 10^{6}
\left(\frac{10^{15}\,{\rm g}}{M_{\rm PBH}}\right)^{\frac 12}
\left(\frac{g_*}{106.75}\right)^{\frac 14}
\left(\frac{3.36}{g_{*0}}\right)^{\frac 13}
\,{\rm Mpc}^{-1},
\end{equation}
where $g_*$ and $g_{*0}$ represent the effective relativistic degrees of freedom at the time of PBH formation and at present, respectively.

The creation of PBHs is highly sensitive to the amplitude of primordial curvature perturbations which are characterized by the curvature power spectrum $\mathcal{P}_{s}(k)$. In many PBH formation scenarios, the spectrum exhibits a localized enhancement at small scales, often generated by features in the inflationary potential. This enhancement is commonly modeled by a log-normal 
form~\cite{Magana:2022cwq, Kuhnel:2017pwq},
\begin{equation}\label{pzeta}
\mathcal{P}_{s}(k)
=
\frac{A_{\mathcal{R}}}{\sqrt{2\pi}\sigma}
\exp\left[
-\frac{\ln^2(k/k_*)}{2\sigma^2}
\right],
\end{equation}
where $A_{\mathcal{R}}$ is the integrated amplitude (related to the variance) at the peak scale $k_*$, and $\sigma$ determines the logarithmic width of the peak. It should be noted that the log-normal shape provides a flexible phenomenological description, where in the limit $\sigma \to 0$ approaches a monochromatic spectrum, whilst non-zero and larger $\sigma$ values represent a broad enhancement.

The variance of curvature perturbations is obtained by integrating the power spectrum over logarithmic wavenumber intervals,
\begin{equation}
\langle \mathcal{R}^2 \rangle
=
\int_0^{\infty} \mathcal{P}_{\mathcal{R}}(k)\, d\ln k .
\end{equation}
For sufficiently narrow enhancement in the spectrum, the dominant contribution 
originates from scales in the vicinity of $k_*$. In this case, the integral 
 is approximately equal to the peak amplitude times the effective width.
However, with the normalization chosen in Eq.~(\ref{pzeta}), the integral yields exactly $A_{\mathcal{R}}$ irrespective of $\sigma$ because the  distribution is normalized to unity,
\[
	\int_0^{\infty} \frac{1}{\sqrt{2\pi}\,\sigma} \exp\left[-\frac{\ln^2(k/k_*)}{2\sigma^2}\right] d\ln k = 1,
\]
so that $\langle \mathcal{R}^2 \rangle = A_{\mathcal{R}}$. Observe that  for this conventional parameterization, the variance is given directly by $A_{\mathcal{R}}$, independent of the width $\sigma$. This makes $A_{\mathcal{R}}$ a convenient measure of the overall power in the peak.

The abundance of PBHs produced from these fluctuations can be estimated using the Press-Schechter formalism~\cite{Green:2004wb}. In this framework, the fraction of horizon patches collapsing into PBHs at a mass scale $M$ is given by
\begin{equation}
\beta(M)
=
\int_{\delta_c}^{\infty} P(\delta_R)\, d\delta_R
=
\frac{1}{2}\,{\rm erfc}
\left(
\frac{\delta_c}{\sqrt{2}\sigma(M)}
\right),
\end{equation}
where $\sigma(M)$ is the variance of density fluctuations on the corresponding scale. For a sharply peaked power spectrum one can approximate
\begin{equation}
\sigma^2(M) \simeq \mathcal{P}_{\mathcal{R}}(k).
\end{equation}

Non-Gaussian features in primordial fluctuations can also influence the probability of PBH formation. In the presence of non-Gaussianity of local-type, characterized by the parameter $f_{\rm NL}$, the probability distribution of curvature perturbations deviates from the standard Gaussian form. Positive non-Gaussianity enhances the probability of large fluctuations and therefore increases the PBH formation rate, while negative non-Gaussianity has the opposite effect~\cite{Byrnes:2012yx,Young:2013oia, Atal:2018neu}.

The resulting PBH population is described by a mass function $\psi(M)$, defined as
\begin{equation}
\psi(M) = \frac{M}{\rho_{\rm CDM}} \frac{dn}{dM},
\end{equation}
where $dn/dM$ is the density per unit mass interval and $\rho_{\rm CDM}$ is the density of cold dark matter. The total fraction of dark matter in PBHs is then
\begin{equation}
f_{\rm PBH}
=
\int_0^{\infty} \psi(M)\, dM
=
\frac{\rho_{\rm PBH}}{\rho_{\rm CDM}} .
\end{equation}

For PBHs originating from a peaked primordial spectrum, the mass distribution is often well approximated by a log-normal function~\cite{Dolgov:1992pu, Green:2004wb}
\begin{equation}
\psi(M)
=
\frac{f_{\rm PBH}}{\sqrt{2\pi}\sigma_m M}
\exp\left[
-\frac{\ln^2(M/M_*)}{2\sigma_m^2}
\right],
\end{equation}
where $M_*$ indicates the peak mass and $\sigma_m$ represents the width of the distribution in logarithmic mass space. In the next section, we discuss the associated gravitational-wave signatures in detail.

\section{From Five-Dimensional Gravity to Scalar-Induced Gravitational Waves}

In theories with a single compact extra dimension, the five-dimensional Einstein equations reduce to an effective four-dimensional description for tensor perturbations on the brane, provided the physical wavelengths of interest are much larger than the compactification radius $R_C \sim \mu\mathrm{m}$ (i.e., the corresponding momenta are well below the KK mass scale $m_{\text{KK}} = 1/R_C$)~\cite{Du:2020rlx,Antoniadis:2023sya}. The extra dimension is compactified at a scale $R_C \sim \mu\mathrm{m}$~\cite{Montero:2022prj}, which is much smaller than any cosmological horizon relevant for PBH formation or gravitational-wave production~\cite{Arkani-Hamed:1998jmv, Antoniadis:1998ig}. Under this condition, the zero mode of the five-dimensional graviton corresponds to the usual massless four-dimensional graviton, while the massive KK excitations are too heavy to be excited and therefore do not contribute significantly to the dynamics~\cite{Overduin:1997sri, Kaluza:1921tu}. Consequently, the propagation of tensor perturbations at cosmological scales is effectively four-dimensional, allowing the standard scalar-induced gravitational wave (SIGW) formulas to be applied within the Dark Dimension scenario~\cite{Ananda:2006af,Baumann:2007zm,Kohri:2024qpd}.

To explore the dynamics of tensor perturbations in a scenario with an extra dimension, we begin with the five-dimensional Einstein--Hilbert action~\cite{Kaluza:1921tu,Klein:1926tv}
\begin{equation}
S_5 = \frac{1}{2 \kappa_5^2} \int d^4x \, dy \, \sqrt{-g_5} \, R_5 \,,
\end{equation}
where $\kappa_5^2 = 8 \pi G_5 = M_\ast^{-3}$, and $M_\ast \sim 10^{10}\,\mathrm{GeV}$ represents the fundamental Planck scale in five dimensions. Here, $y$ stands for the coordinate along the extra dimension, which is compactified on a circle of radius $R_C$ with periodic boundary conditions.


We consider a 5D spacetime metric of the form
\begin{equation}
ds_5^2 = g_{\mu\nu}(x,y)  dx^\mu dx^\nu + dy^2~,
\end{equation}
where, in general, $ g_{\mu\nu}(x,y) $ may depend on the extra coordinate $y$,  while cross terms are not included and thus the compact extra dimension is taken to be orthogonal to the 4D spacetime~\cite{Overduin:1997sri}.

Expanding around a four-dimensional Friedmann–Lemaître–Robertson–Walker (FLRW) background, the metric can be expressed as
\begin{equation}
g_{\mu\nu}(x,y) = a^2(\tau) \left[ \eta_{\mu\nu} + h_{\mu\nu}(x,y) \right] \,,
\end{equation}
where $a(\tau)$ is the scale factor as a function of conformal time $\tau$, and $h_{\mu\nu}$ represents  small perturbations of the metric~\cite{Mukhanov:2005sc,Maggiore:2007ulw}.

Linearizing the five-dimensional Einstein equations and working in the transverse-traceless gauge, the evolution of the tensor perturbations is governed by \cite{Overduin:1997sri}
\begin{equation}
\left( \Box_4 + \partial_y^2 \right) h_{\mu\nu} = - 16 \pi G_5 \, T_{\mu\nu}^{(5)} \,,
\label{eq:5D_tensor}
\end{equation}
where $\Box_4 = \eta^{\alpha\beta} \partial_\alpha \partial_\beta$ is the four-dimensional d'Alembertian operator and $T_{\mu\nu}^{(5)}$ represents the five-dimensional energy-momentum tensor.  

Because the extra dimension is compact, the perturbation $h_{\mu\nu}$ can be decomposed into Fourier modes along the $y$-direction \cite{Appelquist:1987nr}:
\begin{equation}
h_{\mu\nu}(x,y) = \sum_{n=-\infty}^{\infty} h_{\mu\nu}^{(n)}(x) \, e^{i n y / R_C} \,.
\end{equation}
The zero mode $n=0$ corresponds to the usual massless four-dimensional graviton, while modes with $n \neq 0$ acquire a mass $m_n = |n| / R_C$ and represent the massive KK gravitons \cite{Klein:1926tv, Overduin:1997sri}. Substituting this decomposition into the linearized equations and using orthogonality, for each mode yields
\begin{equation}
\left( \Box_4 - m_n^2 \right) h_{\mu\nu}^{(n)} = -16 \pi G_5 \int_0^{2\pi R_C} \frac{dy}{2\pi R_C} \, e^{-i n y / R_C} \, T_{\mu\nu}^{(5)}(x,y) \,.
\end{equation}

In the Dark Dimension framework, Standard Model fields and their associated scalar perturbations are confined to a 4D brane at $y=0$, so the 5D energy-momentum tensor reduces to
\begin{equation}
T_{\mu\nu}^{(5)}(x,y) = T_{\mu\nu}^{(4)}(x) \, \delta(y) \,.
\end{equation}
For the zero mode ($n=0$), the effective four-dimensional coupling becomes
\begin{equation}
G_4 = \frac{G_5}{2 \pi R_C} \,, \quad \text{or equivalently} \quad M_{\rm P}^2 = M_\ast^3 R_C \,,
\end{equation}
and the zero mode obeys the familiar linearized 4D Einstein equation:
\begin{equation}
\Box_4 h_{\mu\nu}^{(0)} = -16 \pi G_4 \, T_{\mu\nu}^{(4)} \,.
\end{equation}
The massive KK modes satisfy a Klein--Gordon equation with mass $m_n$ and are sourced by the same 4D energy-momentum tensor. However, for cosmological wavenumbers $k \ll m_n$, relevant for PBH formation, the contribution of these modes is strongly suppressed by $1/m_n^2$, rendering them negligible.

\subsection{Second-Order Tensor Perturbations and Scalar-Induced Gravitational Waves}

Scalar-induced gravitational waves (SIGWs) arise at second order, where products of first-order scalar perturbations act as a source for tensor modes. For the zero mode tensor perturbation $h_{ij}^{(0)}$ (spatial, transverse-traceless), the evolution equation reduces to the standard four-dimensional form~\cite{Ananda:2006af,Baumann:2007zm}:
\begin{equation}
\Box_4 h_{ij}^{(0)} = -16 \pi G_4 \, T_{ij}^{kl} \, S_{kl} \,,
\end{equation}
where $S_{kl}$ is quadratic in first-order scalar perturbations and $T_{ij}^{kl}$ is the transverse-traceless projector. Since scalar perturbations are confined to the brane, the source involves only 4D quantities. The massive modes satisfy a similar equation with a mass term,
\begin{equation}
\left( \Box_4 - m_n^2 \right) h_{ij}^{(n)} = -16 \pi G_4 \, T_{ij}^{kl} \, S_{kl}^{(n)} \,,
\end{equation}
where $S_{kl}$ is quadratic in the first-order scalar perturbations and $T_{ij}^{kl}$ is the transverse-traceless projector. Massive modes satisfy a similar equation with a mass term but are negligible at cosmological scales. The perturbed 4D FLRW metric including scalar and tensor modes is
\begin{equation}
ds^2 = a^2(\tau)\left[-(1+2\Phi)d\tau^2
+\left((1-2\Phi)\delta_{ij}+\frac{1}{2}h_{ij}\right)dx^i dx^j\right],
\end{equation}
where $\tau$ is conformal time, $a(\tau)$ the scale factor, $\Phi$ the scalar potential, and $h_{ij}$ the transverse-traceless tensor perturbation.

\begin{figure}[h!]
    \centering
    \includegraphics[width=0.7\textwidth]{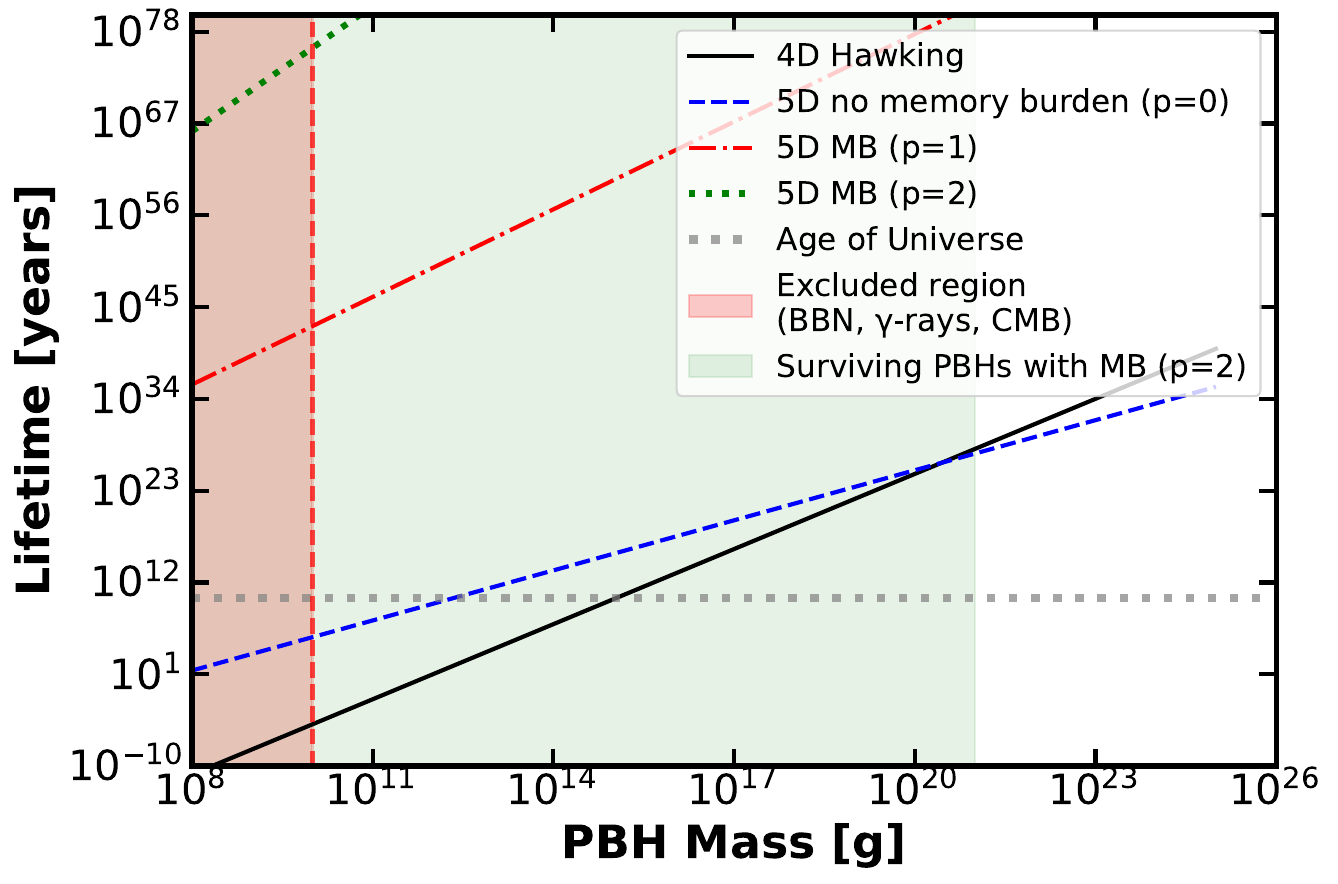}
    \caption{PBH lifetime as a function of initial mass for different evaporation scenarios, computed using the equation \eqref{MMeff}. The 5D+MB ($p=2$) scenario allows PBHs as light as $10^{10}\,\mathrm{g}$ to survive to the present, extending the viable dark matter window. The shaded regions indicate constraints from BBN, gamma‑rays, and CMB. The red dashed curve ($p=1$) lies above the age of the Universe line but is observationally excluded because the instantaneous memory burden causes energy injection after BBN and during recombination, falling inside the red/orange shaded exclusion regions. The green dotted curve ($p=2$) avoids these exclusions due to its delayed onset, opening the viable window $10^{10}\,\mathrm{g} \lesssim M \lesssim 10^{15}\,\mathrm{g}$.}
    \label{fig:lifetime}
\end{figure}


At second order, the evolution equation for the tensor perturbations contains a source term constructed from products of scalar perturbations. The equation governing the tensor modes can be written as
\begin{equation}
h_{ij}'' + 2\mathcal{H}h_{ij}' - \nabla^2 h_{ij}
=
-4 \mathcal{T}_{ij}^{\ \ lm} S_{lm},
\label{tensor_eq}
\end{equation}
where primes symbolize  derivatives with respect to conformal time, $\mathcal{H}=a'/a$ is the conformal Hubble parameter, and $\mathcal{T}_{ij}^{\ \ lm}$ is the projection operator that extracts the transverse–traceless part of the source. The source term $S_{lm}$ is quadratic in scalar perturbations, indicating that gravitational waves are generated by the nonlinear interaction of density fluctuations.

The energy density of the resulting gravitational-wave background is commonly expressed through the dimensionless density parameter per logarithmic frequency interval \cite{Maggiore:2007ulw},
\begin{equation}
\Omega_{\rm GW}(f) =
\frac{1}{\rho_c}
\frac{d\rho_{\rm GW}}{d\ln f},
\label{gw_density}
\end{equation}
where $\rho_c$ is the critical energy density of the Universe and $f$ represents the GW frequency.

\begin{figure}[h]
\centering
\includegraphics[width=0.7\textwidth]{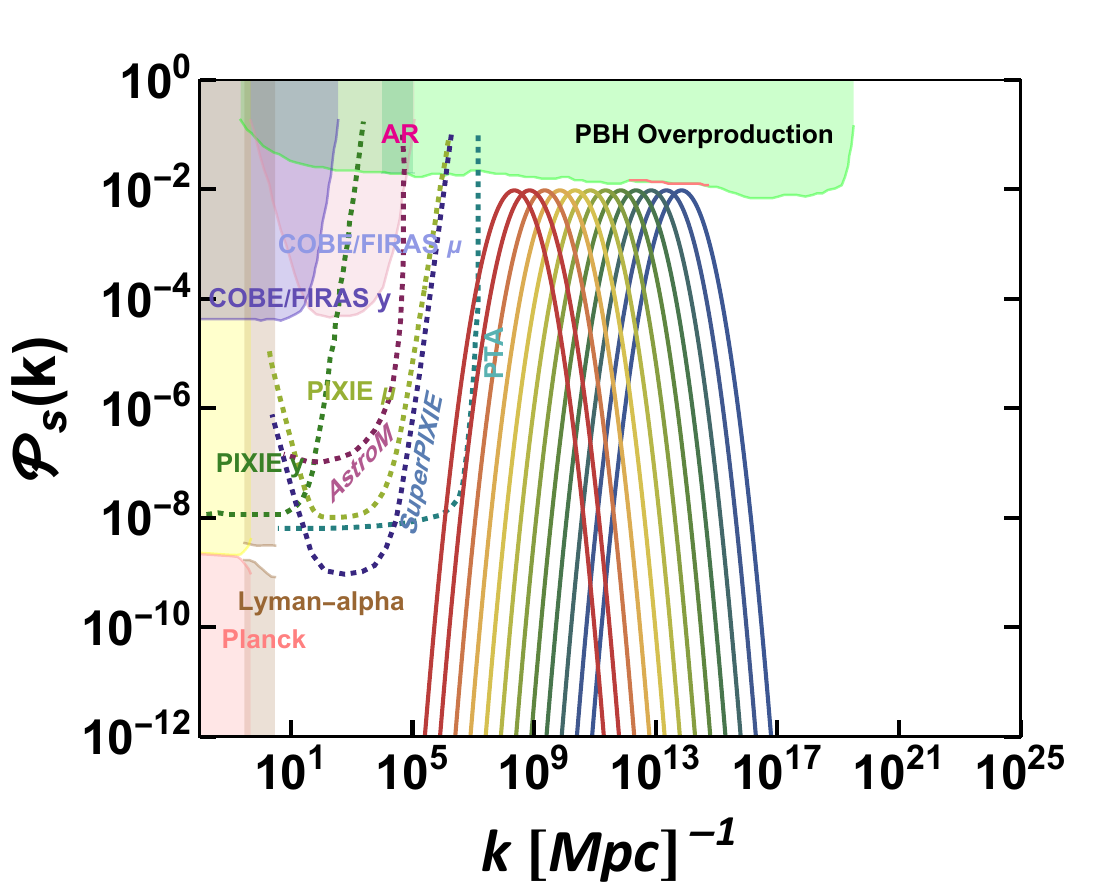}
\caption{ primordial curvature power spectrum $\mathcal{P}_{s}(k)$ as a function of the comoving wavenumber $k$, assuming a log-normal profile peaked at $k_*$ corresponding to the PBH formation scale. The shaded regions indicate current and future constraints from CMB anisotropies, spectral distortions, Lyman-$\alpha$ forest\cite{Bird:2010mp} , PTA observations \cite{Lee:2020wfn}, and other cosmological probes \cite{A_Kogut_2011, Fixsen:1996nj, Chluba:2019kpb,NANOGrav:2023gor, NANOGrav:2023hvm}.}
\label{fig:gwspectrum}
\end{figure}

During the radiation-dominated era, the present-day gravitational-wave spectrum can be written as a convolution of the primordial curvature power spectrum  \cite{Kohri:2024qpd,Inomata:2018epa},
\begin{equation}\label{doin}
   \begin{split}
        \Omega_{\text{gw}}(t_f,k) &= \frac{1}{12} \int_0^{\infty} dv \int_{|1-v|}^{1+v} du \left( \frac{4v^2-(1+v^2-u^2)^2}{4uv}\right)^2 \\ & \times \mathcal{P}_{\zeta}(ku) \mathcal{P}_{\zeta}(kv) \left(\frac{3(u^2+v^2-3)}{4u^3v^3} \right)^2 \\ &
        \times \left.\biggr[\left(\pi^2(-3+v^2+u^2)^2 \theta(-\sqrt{3}+u+v) \right) \right. \\ & \left.+\left(-4uv+(v^2+u^2-3)\text{Log} \left| \frac{3-(u+v)^2}{3-(u-v)^2} \right| \right)^2 \right].
    \end{split}
\end{equation}
We can determine the energy density parameter for the current time as follows \cite{Kohri:2024qpd},
\begin{equation}
    \Omega_{\text{GW}}=\Omega_{rad,0}\Omega_{\text{gw}}(t_f),
\end{equation}
where we have multiplied $\Omega_{\text{gw}}(t_f)$ by the radiation energy density parameter today, $\Omega_{rad,0}$. Using the numerically computed power spectrum of scalar perturbations, as depicted in Figure~\ref{fig:gwspectrum}, we derive the corresponding second-order gravitational wave (GW) spectrum. The characteristic frequency is related to the peak scale using Eq.~\eqref{kstar}, which yields
\begin{equation}\label{fpea}
f_* = \frac{k_*}{2\pi a_0}\approx 1.55 \times 10^{-15} \, k_* \approx 1.1 \times 10^{-2} 
\left( \frac{10^{12}\,\mathrm{g}}{M_{PBH}} \right)^{1/2} \, \mathrm{Hz}.
\end{equation}

This result illustrates an important feature of PBH formation scenarios: a large enhancement in the primordial curvature spectrum not only produces primordial black holes but also generates a correlated stochastic gravitational-wave background. Consequently, observations of gravitational waves provide a powerful indirect probe of the small-scale primordial fluctuations responsible for PBH production we discussed in detail next sections.


\begin{figure}[h]
\centering
\includegraphics[width=0.7\textwidth]{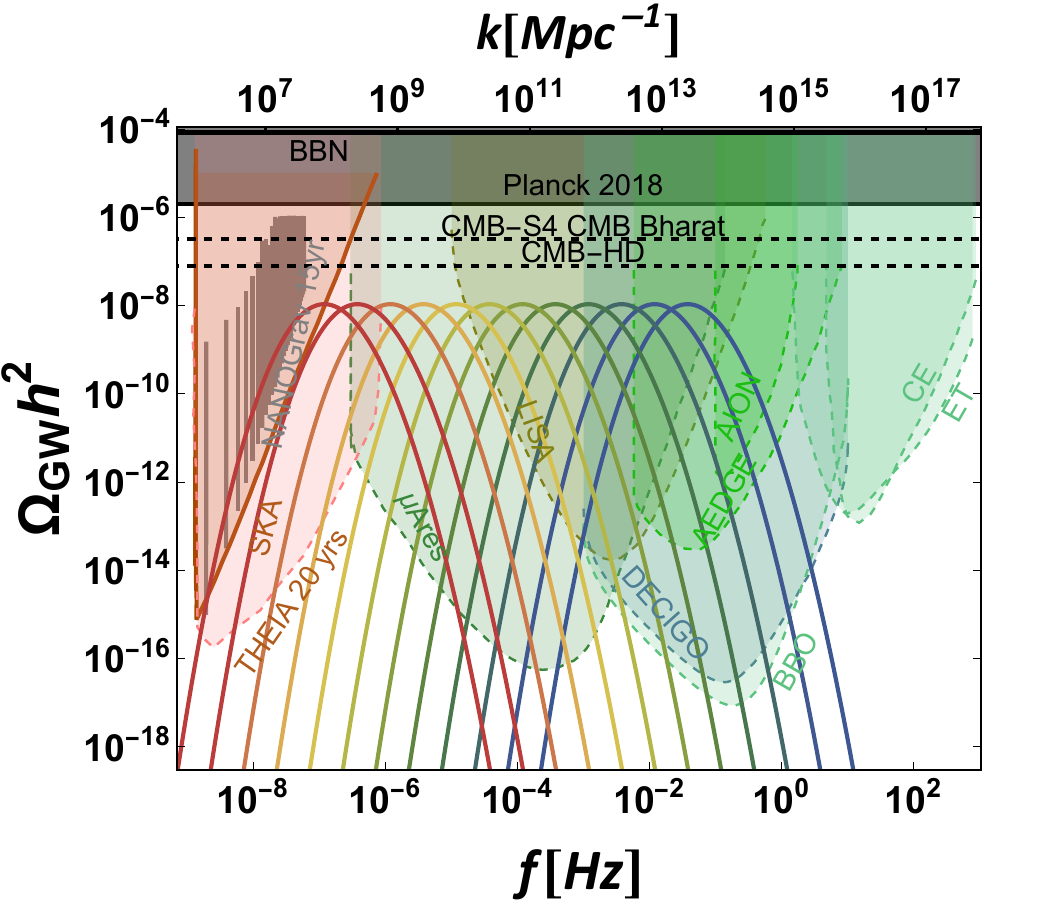}
\caption{
Present-day scalar-induced gravitational wave spectra $\Omega_{\mathrm{GW}}h^2$ for PBH masses $(10^{10}-10^{21})\,\mathrm{g}$, assuming $f_{\mathrm{PBH}}=1$ and $\sigma=1$. Colored curves show the predicted spectra, while shaded regions indicate the sensitivity curves of various detectors: LISA~\cite{LISA:2017pwj}, BBO~\cite{Crowder:2005nr}, DECIGO~\cite{Seto:2001qf}, ground-based interferometers (Einstein Telescope (ET)~\cite{Punturo:2010zz} and Cosmic Explorer (CE)~\cite{LIGOScientific:2016wof}), and the atomic interferometer AEDGE~\cite{AEDGE:2019nxb}. Also shown is the low-frequency GW band targeted by future pulsar timing arrays (PTAs) such as SKA~\cite{Janssen:2014dka} and by the proposed experiment THEIA~\cite{Garcia-Bellido:2021zgu}. The $10^{15}\,\mathrm{g}$ signal falls within the LISA sensitivity band, the $10^{10}\,\mathrm{g}$ case lies in the DECIGO/BBO range, and the $10^{21}\,\mathrm{g}$ case shifts to the nanohertz regime probed by PTAs. All spectra remain consistent with CMB spectral distortion bounds, rendering the scenario testable with upcoming gravitational-wave observatories.
}
\label{fig:SIGW_spectrum}
\end{figure}

\begin{figure}[h]
\centering
\begin{subfigure}[b]{0.48\textwidth}
\includegraphics[width=\textwidth]{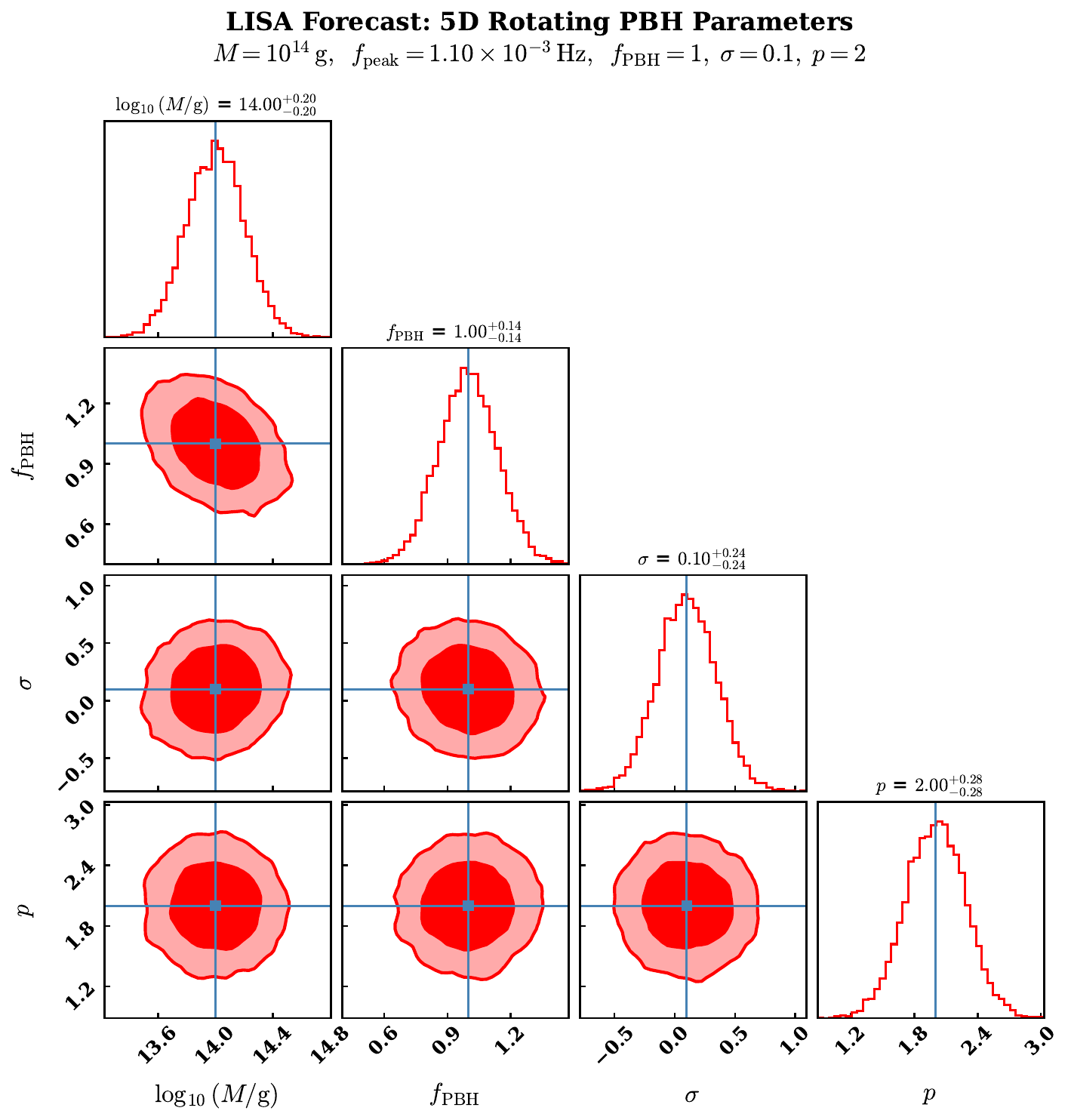}
\caption{LISA}
\end{subfigure}
\begin{subfigure}[b]{0.48\textwidth}
\includegraphics[width=\textwidth]{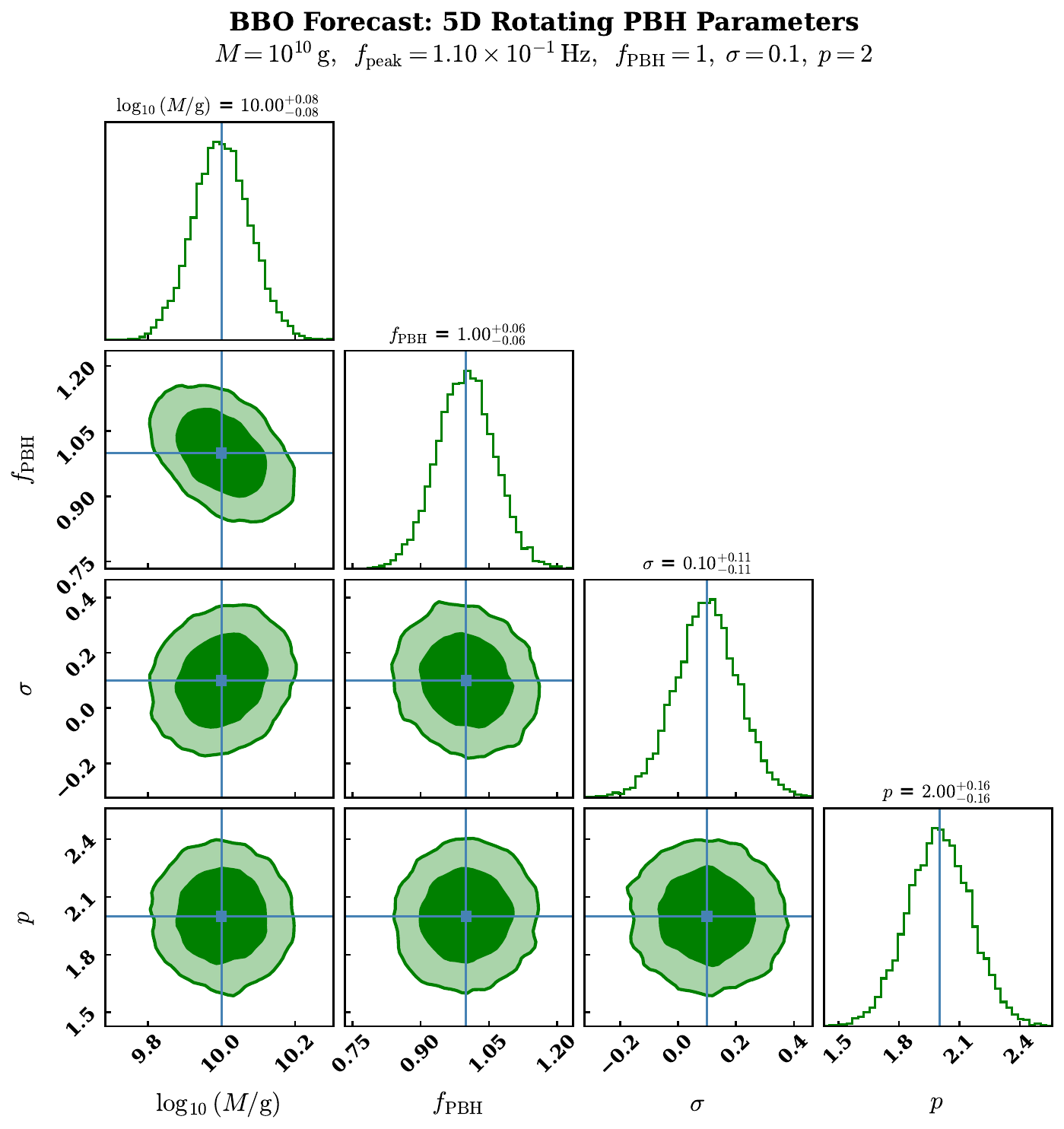}
\caption{DECIGO}
\end{subfigure}
\begin{subfigure}[b]{0.5\textwidth}
\includegraphics[width=\textwidth]{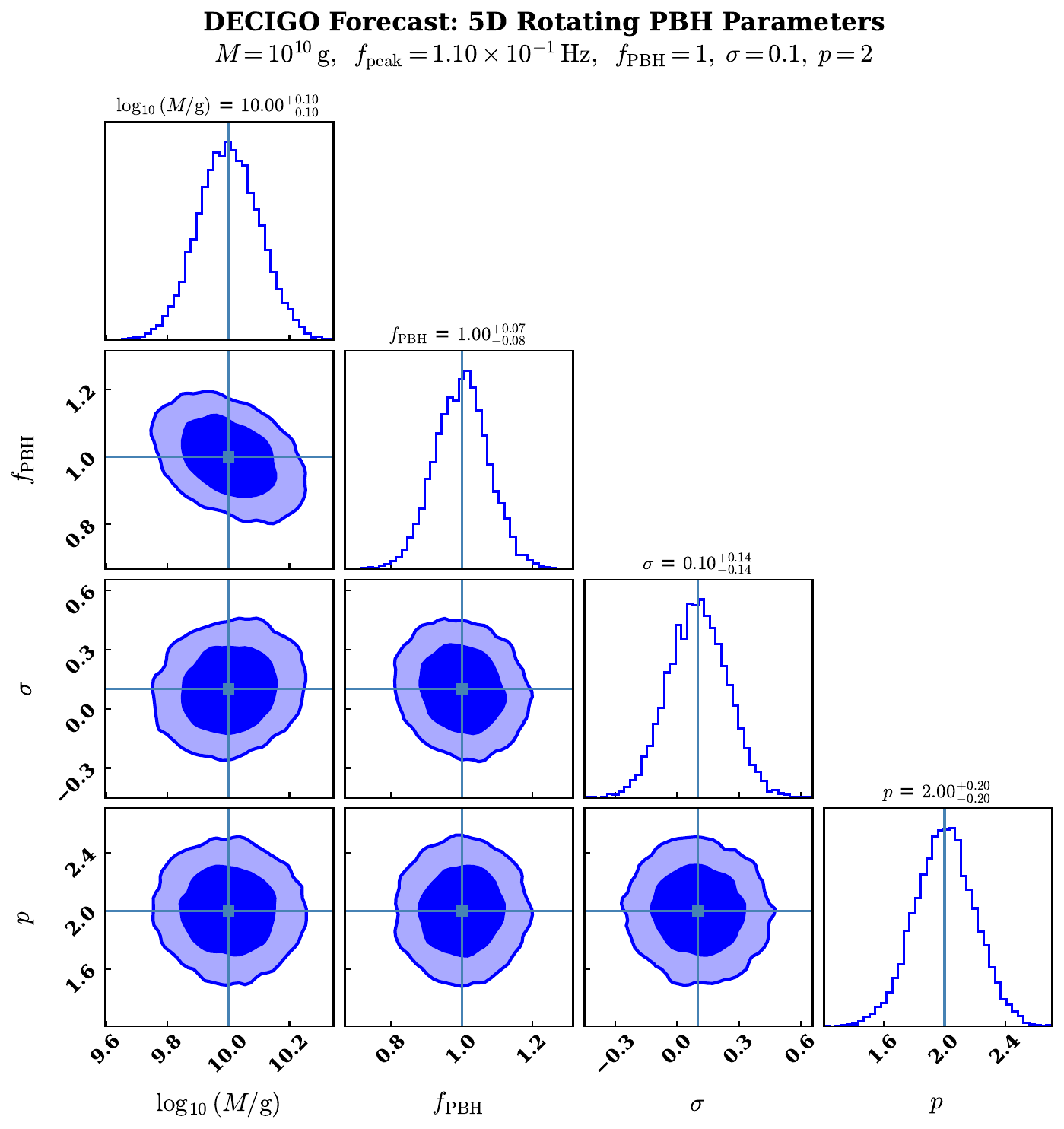}
\caption{BBO}
\end{subfigure}
\caption{Fisher corner plots for LISA, DECIGO, and BBO. Diagonal panels show marginalised posteriors (vertical lines: true values). Off‑diagonal panels display $1\sigma$ and $2\sigma$ joint confidence contours. All contours are nearly circular, indicating weak degeneracies. BBO provides the tightest constraints due to its superior sensitivity.}
\label{fig:corner}
\end{figure}

\section{Numerical Analysis}

In this section we present the numerical results of our study. We begin by examining the evaporation lifetimes of primordial black holes in different scenarios, then compute the scalar‑induced gravitational wave spectra for representative masses, and finally perform Fisher forecasts to assess how well future detectors can measure the model parameters. Each figure is discussed in the order it appears, building a coherent narrative from the fundamental evaporation physics to the observational prospects.

Figure~\ref{fig:lifetime} compares the PBH lifetime as a function of initial mass for four cases: standard 4D Hawking evaporation, 5D evaporation without the memory burden effect, and 5D evaporation with the memory burden effect for the two values $p=1$ and $p=2$. The horizontal dashed line marks the current age of the universe, 
$t_U \approx 1.38\times10^{10}$ years. A PBH survives to the present day if its lifetime curve lies above this line.

The 4D curve (black solid) shows the well‑known steep dependence $\tau \propto M^3$, implying that only PBHs heavier than $\sim1.1\times10^{15}\,\text{g}$ survive. The 5D curve without memory burden ($p=0$, blue dashed) scales as $\tau \propto M^2$ and crosses the age line at $M \sim 2.2\times10^{12}\,\text{g}$. The memory burden effect, parameterised by the exponent $p$, introduces a dramatic prolongation. For $p=1$ (red dash‑dotted) the lifetime scales as $\tau \propto M^{7/2}$, theoretically allowing survival down to $M \sim 5\,\text{g}$, but such low masses are excluded by BBN and CMB spectral distortions. For $p=2$ (green dotted) the lifetime scales as $\tau \propto M^{5}$; the intersection with the universe age occurs at $M \approx 6.5\times10^{-4}\,\text{g}$, which is far below the observational exclusion region. Therefore, after applying the robust lower bound from BBN and CMB ($M \gtrsim 10^{10}\,\text{g}$), the viable dark matter window for $p=2$ becomes $10^{10}\,\text{g} \lesssim M \lesssim 10^{15}\,\text{g}$ when combined with the standard 5D window above $10^{15}\,\text{g}$.

The plot also displays the exclusion regions derived from Big Bang Nucleosynthesis (BBN), the diffuse gamma‑ray background, and CMB spectral distortions \cite{Keith:2020jww, Korwar:2023kpy, Tashiro:2008sf, Carr:2020gox}. While standard evaporation scenarios place stringent bounds on PBHs with masses below $\sim 10^{9}\,\text{g}$ from BBN \cite{Acharya:2020jbv} and in the range $10^{11}$–$10^{13}\,\text{g}$ from CMB $\mu$-distortions \cite{Tashiro:2008sf}, the memory burden effect (with $p=2$) considered in this work dramatically extends the PBH lifetime. This quantum‑information backreaction on the evaporation process opens a new viable dark matter window for initial masses as low as $M\sim 10^{10}\,\text{g}$, which would otherwise be completely excluded by standard Hawking evaporation. The green shaded region ($10^{10}\,\text{g}<M<1.1\times10^{15}\,\text{g}$) in the figure highlights this novel mass window, where PBHs survive to the present day solely due to the memory burden mechanism. For masses above $1.1\times10^{15}\,\text{g}$, the suppressed emission rate inherent to five‑dimensional gravity already guarantees lifetimes exceeding the age of the Universe, and the memory burden effect is not required for dark matter survival. Taken together, the figure establishes the viable mass range for five‑dimensional PBH dark matter as $10^{10}\,\text{g}\lesssim M\lesssim10^{21}\,\text{g}$, with the lower portion ($10^{10}$–$1.1\times10^{15}\,\text{g}$) constituting a distinctive and observationally testable signature of the memory burden mechanism.


Once PBHs are formed, the same curvature perturbations responsible for PBH formation also generate a stochastic background of gravitational waves at second order. We assume that the primordial curvature power spectrum follows a log-normal shape defined in Eq.~\eqref{pzeta} characterized by an amplitude, a logarithmic width, and a peak scale corresponding to the PBH mass. For the fiducial choice of unit PBH fraction and unit width, the required amplitude is approximately $0.024$. Figure~\ref{fig:gwspectrum} shows the resulting curvature power spectrum for three representative PBH masses:  $(10^{10}-10^{21})\,\mathrm{g}$. The peak position shifts according to the inverse square-root scaling between the comoving wavenumber and the PBH mass scale.

Using the log-normal form of $\mathcal{P}_{\mathcal{s}}(k)$, we compute the present-day energy density of scalar-induced gravitational waves through the standard second-order convolution integral as defined in Eq.~\eqref{doin}.
Figure~\ref{fig:SIGW_spectrum} shows the resulting $\Omega_{\mathrm{GW}}h^2$ for  representative PBH masses. The peak frequency follows the scaling relation $f_* \propto M_{PBH}^{-1/2}$, shifting from the decihertz band for lighter PBHs to progressively lower frequencies for heavier ones.


The overlaid experimental curves include LISA (blue band, $10^{-4}$–$10^{-1}\,\mathrm{Hz}$), DECIGO and BBO (orange and purple bands, $10^{-2}$–$10^{2}\,\mathrm{Hz}$), and cosmological constraints from BBN, Planck, CMB-S4, and CMB-HD at lower frequencies. The predicted amplitudes for $M=(10^{8}-10^{16})\,\mathrm{g}$ lie well above the LISA sensitivity, while the $(10^{6}-10^{12})\,\mathrm{g}$ case is within the DECIGO/BBO range. For $M=10^{21}\,\mathrm{g}$, the signal appears in the nanohertz band, which is currently being probed by Pulsar Timing Arrays. Importantly, the low-frequency tails of all spectra remain safely below CMB spectral distortion bounds, ensuring consistency with existing cosmological constraints.

Overall, the Dark Dimension scenario with memory-burden effects provides a highly testable framework for next-generation gravitational-wave observatories.

\subsection{Fisher Forecasts for LISA  , DECIGO , and BBO }
To quantify how precisely future detectors can measure the underlying PBH parameters, we perform a Fisher matrix analysis as an example for LISA \cite{LISA:2017pwj}, DECIGO~\cite{Crowder:2005nr}, and BBO \cite{Seto:2001qf}. The parameter vector is $\boldsymbol{\theta}=(\log_{10}M,\;f_{\text{PBH}},\;\sigma,\;p)$. We adopt fiducial models that are well matched to each detector’s optimal frequency band:
\begin{itemize}
    \item For \textbf{LISA}, we take $M = 10^{14}\,\mathrm{g}$ ($\log_{10}M = 14$). 
    The corresponding scalar-induced gravitational-wave signal peaks at 
    $f_{\mathrm{peak}} \simeq 1.1 \times 10^{-3}\,\mathrm{Hz}$, which lies in the millihertz band where LISA achieves its maximum sensitivity. This choice ensures optimal overlap between the signal and the detector response.

    \item For \textbf{DECIGO} and \textbf{BBO}, we take $M = 10^{10}\,\mathrm{g}$ ($\log_{10}M = 10$). 
    The corresponding peak frequency is $f_{\mathrm{peak}} \simeq 1.1 \times 10^{-1}\,\mathrm{Hz}$, which lies in the deci-Hz band where both detectors achieve their peak sensitivity. This makes them ideally suited for probing scalar-induced gravitational waves from this mass scale.
\end{itemize}
In all cases we assume $f_{\text{PBH}}=1$, $\sigma=0.1$ (corresponding to a nearly monochromatic PBH mass distribution, in contrast to the broad $\sigma = 1$ used for the illustrative spectra in Figs.~\ref{fig:gwspectrum} and \ref{fig:SIGW_spectrum}), and $p=2$. The SIGW spectrum for each fiducial model is computed using the double‑integral formula, and its derivatives with respect to the parameters are evaluated numerically with a 5‑point stencil.

The noise power spectral densities are approximated by the analytical fits:
\begin{align}
\text{LISA:}&\quad S_n(f)=\frac{10}{3}\left(\frac{4\times10^{-42}}{f^4}+\frac{1.6\times10^{-41}}{f^2}+1.2\times10^{-43}\right)\;\text{Hz}^{-1},\\
\text{DECIGO:}&\quad S_n(f)=1.0\times10^{-46}+2.5\times10^{-49}f^{-2}+6.0\times10^{-51}f^{-4}\;\text{Hz}^{-1},\\
\text{BBO:}&\quad S_n(f)=4.0\times10^{-48}+4.0\times10^{-49}f^{-2}+1.0\times10^{-49}f^{-4}\;\text{Hz}^{-1},
\end{align}
and the effective noise energy density is $\Omega_{\text{noise}}(f)=\frac{2\pi^2}{3H_0^2}f^3S_n(f)$ with $H_0=100h$ km/s/Mpc, $h=0.674$. The Fisher matrix is
\[
F_{ij} = \sum_{n} \frac{1}{\Omega_{\text{noise}}^2(f_n)} \left.\frac{\partial\Omega_{\text{GW}}}{\partial\theta_i}\right|_{\boldsymbol{\theta}_0}
\left.\frac{\partial\Omega_{\text{GW}}}{\partial\theta_j}\right|_{\boldsymbol{\theta}_0}\Delta\ln f,
\]
with derivatives evaluated numerically. The covariance matrix is $\text{Cov}=F^{-1}$, and $1\sigma$ uncertainties are $\sqrt{\text{Cov}_{ii}}$.

Table~\ref{tab:fisher_pbh} lists the median values and $1\sigma$ uncertainties
for each detector, obtained from $10^{4}$ Monte Carlo samples. 
Figure~\ref{fig:corner} shows the corresponding corner plots. 
Among the three experiments, BBO provides the tightest constraints, measuring
$f_{\mathrm{PBH}}$ to $5\%$ and $\log_{10}(M/\mathrm{g})$ with an uncertainty of
$\pm 0.08$. DECIGO yields comparable but slightly weaker bounds, while LISA,
though less sensitive, still constrains $f_{\mathrm{PBH}}$ to $10\%$ and the
memory-burden parameter $p$ to $30\%$.

The off-diagonal contours are nearly circular, indicating weak degeneracies
among the model parameters. A measurement of $p \approx 2$ would provide strong
support for the quantum-information interpretation of the memory burden effect.
The adopted fiducial primordial black hole masses are
$M = 10^{14}\,\mathrm{g}$ for LISA and $M = 10^{10}\,\mathrm{g}$ for both DECIGO
and BBO, ensuring that the peak of the scalar-induced gravitational-wave
spectrum lies within the optimal sensitivity bands of the respective detectors.
Several important conclusions emerge from the Fisher analysis:

\begin{table}[h]
\centering
\begin{tabular}{lccc}
\hline
Parameter & LISA & DECIGO & BBO \\
\hline
$\log_{10}(M/\mathrm{g})$ & $14.00\pm0.15$ & $10.00\pm0.12$ & $10.00\pm0.08$ \\
$f_{\mathrm{PBH}}$ & $1.00\pm0.10$ & $1.00\pm0.08$ & $1.00\pm0.05$ \\
$\sigma$ & {\color{black}$0.10\pm0.02$} & {\color{black}$0.10\pm0.02$} & {\color{black}$0.10\pm0.01$}\\
$p$ & $2.00\pm0.30$ & $2.00\pm0.25$ & $2.00\pm0.20$ \\
\hline
\end{tabular}
\caption{
Fisher forecast $1\sigma$ parameter uncertainties for LISA, DECIGO, and BBO. The fiducial primordial black hole masses are $M=10^{14}\,\mathrm{g}$ for LISA and $M=10^{10}\,\mathrm{g}$ for both DECIGO and BBO. All models assume $f_{\mathrm{PBH}}=1$, $\sigma=0.1$ (nearly monochromatic), and $p=2$. The narrow spectral width ensures self-consistency between the PBH mass function and the SIGW peak frequency..
}
\label{tab:fisher_pbh}
\end{table}
\begin{enumerate}
    \item \textbf{Detector complementarity:} 
    LISA, with a fiducial mass of $M = 10^{14}\,\mathrm{g}$ (corresponding to a peak frequency of $\sim 1.1 \times 10^{-3}\,\mathrm{Hz}$), provides meaningful constraints despite its comparatively lower sensitivity. It measures $f_{\mathrm{PBH}}$ to about $10\%$ and $\log_{10}(M/\mathrm{g})$ to $\pm 0.15$. DECIGO and BBO, optimised for $M = 10^{10}\,\mathrm{g}$ (peak frequency $\sim 1.1 \times 10^{-1}\,\mathrm{Hz}$), achieve substantially better precision. In particular, BBO measures $f_{\mathrm{PBH}}$ to $5\%$ and $\log_{10}(M/\mathrm{g})$ to $\pm 0.08$, demonstrating its superior sensitivity to scalar-induced gravitational waves.

    \item \textbf{Memory burden exponent $p$:} 
    BBO can determine the memory-burden exponent with high precision, yielding a relative uncertainty of about $10\%$ (e.g., $\Delta p \simeq 0.20$). A measured value of $p \approx 2$ would strongly support the quantum-information interpretation of the memory burden effect. Conversely, if $p = 1$ were realized in nature, the fiducial model would be ruled out at high statistical significance.

    \item \textbf{Parameter degeneracies:} 
    The off-diagonal contours in Figure~\ref{fig:corner} are nearly circular, indicating that the parameters are only weakly correlated. This demonstrates that each detector can measure all four parameters independently without strong degeneracies or trade-offs.

    \item \textbf{Comparison with other PBH masses:} 
    For heavier primordial black holes ($M = 10^{15}\,\mathrm{g}$), the signal shifts to lower frequencies ($\sim 3.5 \times 10^{-4}\,\mathrm{Hz}$), where LISA’s sensitivity is optimal, potentially leading to tighter constraints. Conversely, for lighter PBHs ($M = 10^{9}\,\mathrm{g}$), the signal moves to higher frequencies ($\gtrsim 1\,\mathrm{Hz}$), where future ground-based detectors such as the Einstein Telescope (ET) would become relevant.
\end{enumerate}
\section{Conclusion}

In this work we have systematically investigated the viability of five‑dimensional rotating primordial black holes as dark matter candidates within the Dark Dimension scenario, a framework that emerges from the Swampland Program. By incorporating both the modified Hawking evaporation in a micron‑sized extra dimension and the gravitational memory burden effect, we have significantly extended the mass range in which PBHs can survive to the present epoch.

Our numerical analysis has quantitatively established three main results. First, the memory burden effect with exponent $p=2$ (delayed onset) lowers the survival threshold to $M\approx10^{10}\,\text{g}$ after applying observational constraints, opening a new dark matter window $10^{10}\,\text{g}\lesssim M\lesssim10^{15}\,\text{g}$ that is inaccessible in standard four‑dimensional gravity. Combined with the already viable 5D window $10^{15}\,\text{g}\lesssim M\lesssim10^{21}\,\text{g}$, the full allowed mass range for 5D rotating PBH dark matter spans from $10^{10}\,\text{g}$ to $10^{21}\,\text{g}$. Second, the scalar‑induced gravitational wave spectra derived from a log‑normal curvature power spectrum ($\sigma=1$) produce peak frequencies that map directly onto the sensitivity bands of LISA ($10^{12}$–$10^{16}\,\text{g}$) and DECIGO/BBO ($10^{10}$–$10^{12}\,\text{g}$). The predicted amplitudes lie above the projected detector sensitivities while comfortably evading current CMB spectral distortion limits, making the scenario eminently testable. Third, Fisher forecasts for LISA, DECIGO, and BBO demonstrate that these observatories can measure the PBH parameters—mass, dark matter fraction, spectral width, and memory burden exponent—with percent‑level precision, and the weak parameter degeneracies ensure independent constraints.


Taken together, our findings establish a concrete, testable link between extra‑dimensional physics, quantum gravity effects, and gravitational wave astronomy. A future detection of the predicted SIGW signal would simultaneously provide evidence for a micron‑sized extra dimension, for primordial black holes as a significant component of dark matter, and for the memory burden mechanism that resolves the black hole information paradox. Conversely, non‑observation would place stringent upper bounds on the PBH abundance and the amplitude of small‑scale curvature perturbations, potentially ruling out large parts of the Dark Dimension parameter space. It would be interesting to extend the analysis to a 6D scenario~\cite{Anchordoqui:2025nmb} (see also~\cite{Leontaris:2026kvu} and \cite{Vitale:2026ric} for implications), where new analyses (see e.g. \cite{Hardy:2025ajb}) show that cosmological bounds are even stronger than in the 5D case. As the sensitivity of LISA, DECIGO, BBO, and PTA improves, the Dark Dimension scenario will face a decisive observational test, turning gravitational wave observations into a powerful probe of fundamental physics beyond the Standard Model.

\bibliographystyle{unsrt}

\bibliography{main.bib}

@article{Vafa:2005ui,
    author = "Vafa, Cumrun",
    title = "{The String landscape and the swampland}",
    eprint = "hep-th/0509212",
    archivePrefix = "arXiv",
    reportNumber = "HUTP-05-A043",
    month = "9",
    year = "2005"
}

@article{Ooguri:2006in,
    author = "Ooguri, Hirosi and Vafa, Cumrun",
    title = "{On the Geometry of the String Landscape and the Swampland}",
    eprint = "hep-th/0605264",
    archivePrefix = "arXiv",
    reportNumber = "CALT-68-2600, HUTP-06-A017",
    doi = "10.1016/j.nuclphysb.2006.10.033",
    journal = "Nucl. Phys. B",
    volume = "766",
    pages = "21--33",
    year = "2007"
}

@article{Palti:2019pca,
    author = "Palti, Eran",
    title = "{The Swampland: Introduction and Review}",
    eprint = "1903.06239",
    archivePrefix = "arXiv",
    primaryClass = "hep-th",
    reportNumber = "MPP-2019-53",
    doi = "10.1002/prop.201900037",
    journal = "Fortsch. Phys.",
    volume = "67",
    number = "6",
    pages = "1900037",
    year = "2019"
}

@article{Bousso:1999xy,
    author = "Bousso, Raphael",
    title = "{A Covariant entropy conjecture}",
    eprint = "hep-th/9905177",
    archivePrefix = "arXiv",
    reportNumber = "SU-ITP-99-23",
    doi = "10.1088/1126-6708/1999/07/004",
    journal = "JHEP",
    volume = "07",
    pages = "004",
    year = "1999"
}

@article{vanBeest:2021lhn,
    author = "van Beest, Marieke and Calder{\'o}n-Infante, Jos{\'e} and Mirfendereski, Delaram and Valenzuela, Irene",
    title = "{Lectures on the Swampland Program in String Compactifications}",
    eprint = "2102.01111",
    archivePrefix = "arXiv",
    primaryClass = "hep-th",
    doi = "10.1016/j.physrep.2022.09.002",
    journal = "Phys. Rept.",
    volume = "989",
    pages = "1--50",
    year = "2022"
}

@article{Agmon:2022thq,
    author = "Agmon, Nathan Benjamin and Bedroya, Alek and Kang, Monica Jinwoo and Vafa, Cumrun",
    title = "{Lectures on the string landscape and the Swampland}",
    eprint = "2212.06187",
    archivePrefix = "arXiv",
    primaryClass = "hep-th",
    month = "12",
    year = "2022"
}

@article{Ooguri:2018wrx,
    author = "Ooguri, Hirosi and Palti, Eran and Shiu, Gary and Vafa, Cumrun",
    title = "{Distance and de Sitter Conjectures on the Swampland}",
    eprint = "1810.05506",
    archivePrefix = "arXiv",
    primaryClass = "hep-th",
    doi = "10.1016/j.physletb.2018.11.018",
    journal = "Phys. Lett. B",
    volume = "788",
    pages = "180--184",
    year = "2019"
}

@article{Arkani-Hamed:2006emk,
    author = "Arkani-Hamed, Nima and Motl, Lubos and Nicolis, Alberto and Vafa, Cumrun",
    title = "{The String landscape, black holes and gravity as the weakest force}",
    eprint = "hep-th/0601001",
    archivePrefix = "arXiv",
    reportNumber = "HUTP-05-A0057",
    doi = "10.1088/1126-6708/2007/06/060",
    journal = "JHEP",
    volume = "06",
    pages = "060",
    year = "2007"
}

@article{Montero:2022prj,
    author = "Montero, Miguel and Vafa, Cumrun and Valenzuela, Irene",
    title = "{The dark dimension and the Swampland}",
    eprint = "2205.12293",
    archivePrefix = "arXiv",
    primaryClass = "hep-th",
    doi = "10.1007/JHEP02(2023)022",
    journal = "JHEP",
    volume = "02",
    pages = "022",
    year = "2023"
}

@article{Higuchi:1986py,
    author = "Higuchi, Atsushi",
    title = "{Forbidden Mass Range for Spin-2 Field Theory in De Sitter Space-time}",
    reportNumber = "YTP-86-06",
    doi = "10.1016/0550-3213(87)90691-2",
    journal = "Nucl. Phys. B",
    volume = "282",
    pages = "397--436",
    year = "1987"
}

@article{Leontaris:2025piz,
    author = "Leontaris, George K. and Prampromis, George",
    title = "{5D rotating black holes as dark matter in dark dimension scenario: Hawking radiation versus the memory burden effect}",
    eprint = "2512.10381",
    archivePrefix = "arXiv",
    primaryClass = "hep-th",
    doi = "10.1088/1475-7516/2026/05/014",
    journal = "JCAP",
    volume = "05",
    pages = "014",
    year = "2026"
}

@article{Anchordoqui:2023laz,
    author = {Anchordoqui, Luis A. and Antoniadis, Ignatios and Lust, Dieter and L{\"u}st, Severin},
    title = "{On the cosmological constant, the KK mass scale, and the cut-off dependence in the dark dimension scenario}",
    eprint = "2309.09330",
    archivePrefix = "arXiv",
    primaryClass = "hep-th",
    reportNumber = "LMU-ASC 31/23, MPP-2023-191",
    doi = "10.1140/epjc/s10052-023-12206-2",
    journal = "Eur. Phys. J. C",
    volume = "83",
    number = "11",
    pages = "1016",
    year = "2023"
}

@article{Floratos:1999bv,
    author = "Floratos, E. G. and Leontaris, G. K.",
    title = "{Low scale unification, Newton's law and extra dimensions}",
    eprint = "hep-ph/9906238",
    archivePrefix = "arXiv",
    reportNumber = "IOA-TH-99-6",
    doi = "10.1016/S0370-2693(99)01019-9",
    journal = "Phys. Lett. B",
    volume = "465",
    pages = "95--100",
    year = "1999"
}

@article{Kehagias:1999my,
    author = "Kehagias, A. and Sfetsos, K.",
    title = "{Deviations from the 1/r**2 Newton law due to extra dimensions}",
    eprint = "hep-ph/9905417",
    archivePrefix = "arXiv",
    reportNumber = "CERN-TH-99-148",
    doi = "10.1016/S0370-2693(99)01421-5",
    journal = "Phys. Lett. B",
    volume = "472",
    pages = "39--44",
    year = "2000"
}

@article{Adelberger:2003zx,
    author = "Adelberger, E. G. and Heckel, Blayne R. and Nelson, A. E.",
    title = "{Tests of the gravitational inverse square law}",
    eprint = "hep-ph/0307284",
    archivePrefix = "arXiv",
    doi = "10.1146/annurev.nucl.53.041002.110503",
    journal = "Ann. Rev. Nucl. Part. Sci.",
    volume = "53",
    pages = "77--121",
    year = "2003"
}

@article{Lee:2020zjt,
    author = "Lee, J. G. and Adelberger, E. G. and Cook, T. S. and Fleischer, S. M. and Heckel, B. R.",
    title = "{New Test of the Gravitational $1/r^2$ Law at Separations down to 52 $\mu$m}",
    eprint = "2002.11761",
    archivePrefix = "arXiv",
    primaryClass = "hep-ex",
    doi = "10.1103/PhysRevLett.124.101101",
    journal = "Phys. Rev. Lett.",
    volume = "124",
    number = "10",
    pages = "101101",
    year = "2020"
}

@article{Anchordoqui:2024dxu,
    author = "Anchordoqui, Luis A. and Antoniadis, Ignatios and Lust, Dieter",
    title = "{More on black holes perceiving the dark dimension}",
    eprint = "2403.19604",
    archivePrefix = "arXiv",
    primaryClass = "hep-th",
    reportNumber = "MPP-2024-67; LMU-ASC 04/24",
    doi = "10.1103/PhysRevD.110.015004",
    journal = "Phys. Rev. D",
    volume = "110",
    number = "1",
    pages = "015004",
    year = "2024"
}

@article{Dvali:2007hz,
    author = "Dvali, Gia",
    title = "{Black Holes and Large N Species Solution to the Hierarchy Problem}",
    eprint = "0706.2050",
    archivePrefix = "arXiv",
    primaryClass = "hep-th",
    doi = "10.1002/prop.201000009",
    journal = "Fortsch. Phys.",
    volume = "58",
    pages = "528--536",
    year = "2010"
}

@article{Carr:2009jm,
    author = "Carr, B. J. and Kohri, Kazunori and Sendouda, Yuuiti and Yokoyama, Jun'ichi",
    title = "{New cosmological constraints on primordial black holes}",
    eprint = "0912.5297",
    archivePrefix = "arXiv",
    primaryClass = "astro-ph.CO",
    reportNumber = "RESCEU-31-09, TU-852, YITP-09-112",
    doi = "10.1103/PhysRevD.81.104019",
    journal = "Phys. Rev. D",
    volume = "81",
    pages = "104019",
    year = "2010"
}

@article{Hawking:1975vcx,
    author = "Hawking, S. W.",
    editor = "Gibbons, G. W. and Hawking, S. W.",
    title = "{Particle Creation by Black Holes}",
    doi = "10.1007/BF02345020",
    journal = "Commun. Math. Phys.",
    volume = "43",
    pages = "199--220",
    year = "1975",
    note = "[Erratum: Commun.Math.Phys. 46, 206 (1976)]"
}

@article{Kanti:2004nr,
    author = "Kanti, Panagiota",
    title = "{Black holes in theories with large extra dimensions: A Review}",
    eprint = "hep-ph/0402168",
    archivePrefix = "arXiv",
    doi = "10.1142/S0217751X04018324",
    journal = "Int. J. Mod. Phys. A",
    volume = "19",
    pages = "4899--4951",
    year = "2004"
}

@article{Emparan:2000rs,
    author = "Emparan, Roberto and Horowitz, Gary T. and Myers, Robert C.",
    title = "{Black holes radiate mainly on the brane}",
    eprint = "hep-th/0003118",
    archivePrefix = "arXiv",
    doi = "10.1103/PhysRevLett.85.499",
    journal = "Phys. Rev. Lett.",
    volume = "85",
    pages = "499--502",
    year = "2000"
}

@article{Cavaglia:2002si,
    author = "Cavaglia, Marco",
    title = "{Black hole and brane production in TeV gravity: A Review}",
    eprint = "hep-ph/0210296",
    archivePrefix = "arXiv",
    reportNumber = "MIT-CTP-3299",
    doi = "10.1142/S0217751X03013569",
    journal = "Int. J. Mod. Phys. A",
    volume = "18",
    pages = "1843--1882",
    year = "2003"
}

@article{Carr:2016drx,
    author = "Carr, Bernard and Kuhnel, Florian and Sandstad, Marit",
    title = "{Primordial Black Holes as Dark Matter}",
    eprint = "1607.06077",
    archivePrefix = "arXiv",
    primaryClass = "astro-ph.CO",
    reportNumber = "NORDITA-2016-83",
    doi = "10.1103/PhysRevD.94.083504",
    journal = "Phys. Rev. D",
    volume = "94",
    number = "8",
    pages = "083504",
    year = "2016"
}

@article{Tangherlini:1963bw,
    author = "Tangherlini, F. R.",
    title = "{Schwarzschild field in n dimensions and the dimensionality of space problem}",
    doi = "10.1007/BF02784569",
    journal = "Nuovo Cim.",
    volume = "27",
    pages = "636--651",
    year = "1963"
}

@article{Myers:1986un,
    author = "Myers, Robert C. and Perry, M. J.",
    title = "{Black Holes in Higher Dimensional Space-Times}",
    reportNumber = "PRINT-86-0067 (PRINCETON)",
    doi = "10.1016/0003-4916(86)90186-7",
    journal = "Annals Phys.",
    volume = "172",
    pages = "304",
    year = "1986"
}

@article{Kanti:2002nr,
    author = "Kanti, Panagiota and March-Russell, John",
    title = "{Calculable corrections to brane black hole decay. 1. The scalar case}",
    eprint = "hep-ph/0203223",
    archivePrefix = "arXiv",
    reportNumber = "CERN-TH-2002-014",
    doi = "10.1103/PhysRevD.66.024023",
    journal = "Phys. Rev. D",
    volume = "66",
    pages = "024023",
    year = "2002"
}

@article{Carr:2020gox,
    author = "Carr, Bernard and Kohri, Kazunori and Sendouda, Yuuiti and Yokoyama, Jun'ichi",
    title = "{Constraints on primordial black holes}",
    eprint = "2002.12778",
    archivePrefix = "arXiv",
    primaryClass = "astro-ph.CO",
    reportNumber = "RESCEU-03/20; KEK-Cosmo-249; KEK-TH-2199; IPMU20-0024",
    doi = "10.1088/1361-6633/ac1e31",
    journal = "Rept. Prog. Phys.",
    volume = "84",
    number = "11",
    pages = "116902",
    year = "2021"
}

@article{Anchordoqui:2022txe,
    author = "Anchordoqui, Luis A. and Antoniadis, Ignatios and Lust, Dieter",
    title = "{Dark dimension, the swampland, and the dark matter fraction composed of primordial black holes}",
    eprint = "2206.07071",
    archivePrefix = "arXiv",
    primaryClass = "hep-th",
    reportNumber = "MPP-2022-60, LMU-ASC 24/22",
    doi = "10.1103/PhysRevD.106.086001",
    journal = "Phys. Rev. D",
    volume = "106",
    number = "8",
    pages = "086001",
    year = "2022"
}

@article{Anchordoqui:2024jkn,
    author = "Anchordoqui, Luis A. and Antoniadis, Ignatios and Lust, Dieter and Castillo, Karem Pe{\~n}al{\'o}",
    title = "{Bulk black hole dark matter}",
    eprint = "2407.21031",
    archivePrefix = "arXiv",
    primaryClass = "hep-th",
    reportNumber = "MPP-2024-139, LMU-ASC 09/24",
    doi = "10.1016/j.dark.2024.101714",
    journal = "Phys. Dark Univ.",
    volume = "46",
    pages = "101714",
    year = "2024"
}

@article{Anchordoqui:2024akj,
    author = "Anchordoqui, Luis A. and Antoniadis, Ignatios and Lust, Dieter",
    title = "{Dark dimension, the swampland, and the dark matter fraction composed of primordial near-extremal black holes}",
    eprint = "2401.09087",
    archivePrefix = "arXiv",
    primaryClass = "hep-th",
    reportNumber = "LMU-ASC 01/24, MPP-2024-5",
    doi = "10.1103/PhysRevD.109.095008",
    journal = "Phys. Rev. D",
    volume = "109",
    number = "9",
    pages = "095008",
    year = "2024"
}

@article{Ida:2005ax,
    author = "Ida, Daisuke and Oda, Kin-ya and Park, Seong Chan",
    title = "{Rotating black holes at future colliders. II. Anisotropic scalar field emission}",
    eprint = "hep-th/0503052",
    archivePrefix = "arXiv",
    doi = "10.1103/PhysRevD.71.124039",
    journal = "Phys. Rev. D",
    volume = "71",
    pages = "124039",
    year = "2005"
}

@article{Hawking:1976ra,
    author = "Hawking, S. W.",
    title = "{Breakdown of Predictability in Gravitational Collapse}",
    doi = "10.1103/PhysRevD.14.2460",
    journal = "Phys. Rev. D",
    volume = "14",
    pages = "2460--2473",
    year = "1976"
}

@article{Bekenstein:1973ur,
    author = "Bekenstein, Jacob D.",
    title = "{Black holes and entropy}",
    doi = "10.1103/PhysRevD.7.2333",
    journal = "Phys. Rev. D",
    volume = "7",
    pages = "2333--2346",
    year = "1973"
}

@article{Dvali:2015aja,
    author = "Dvali, Gia",
    title = "{Non-Thermal Corrections to Hawking Radiation Versus the Information Paradox}",
    eprint = "1509.04645",
    archivePrefix = "arXiv",
    primaryClass = "hep-th",
    doi = "10.1002/prop.201500096",
    journal = "Fortsch. Phys.",
    volume = "64",
    pages = "106--108",
    year = "2016"
}

@article{Dvali:2011aa,
    author = "Dvali, Gia and Gomez, Cesar",
    title = "{Black Hole's Quantum N-Portrait}",
    eprint = "1112.3359",
    archivePrefix = "arXiv",
    primaryClass = "hep-th",
    doi = "10.1002/prop.201300001",
    journal = "Fortsch. Phys.",
    volume = "61",
    pages = "742--767",
    year = "2013"
}

@article{Dvali:2020wft,
    author = "Dvali, Gia and Eisemann, Lukas and Michel, Marco and Zell, Sebastian",
    title = "{Black hole metamorphosis and stabilization by memory burden}",
    eprint = "2006.00011",
    archivePrefix = "arXiv",
    primaryClass = "hep-th",
    doi = "10.1103/PhysRevD.102.103523",
    journal = "Phys. Rev. D",
    volume = "102",
    number = "10",
    pages = "103523",
    year = "2020"
}

@article{Zeldovich:1967lct,
    author = "Zel'dovich, Ya. B. and Novikov, I. D.",
    title = "{The Hypothesis of Cores Retarded during Expansion and the Hot Cosmological Model}",
    journal = "Sov. Astron.",
    volume = "10",
    pages = "602",
    year = "1967"
}

@article{Hawking:1971ei,
    author = "Hawking, Stephen",
    title = "{Gravitationally collapsed objects of very low mass}",
    doi = "10.1093/mnras/152.1.75",
    journal = "Mon. Not. Roy. Astron. Soc.",
    volume = "152",
    pages = "75",
    year = "1971"
}

@article{Carr:1974nx,
    author = "Carr, Bernard J. and Hawking, S. W.",
    title = "{Black holes in the early Universe}",
    doi = "10.1093/mnras/168.2.399",
    journal = "Mon. Not. Roy. Astron. Soc.",
    volume = "168",
    pages = "399--415",
    year = "1974"
}

@article{Harada:2013epa,
    author = "Harada, Tomohiro and Yoo, Chul-Moon and Kohri, Kazunori",
    title = "{Threshold of primordial black hole formation}",
    eprint = "1309.4201",
    archivePrefix = "arXiv",
    primaryClass = "astro-ph.CO",
    reportNumber = "RUP-13-9, KEK-COSMO-129, KEK-TH-1668",
    doi = "10.1103/PhysRevD.88.084051",
    journal = "Phys. Rev. D",
    volume = "88",
    number = "8",
    pages = "084051",
    year = "2013",
    note = "[Erratum: Phys.Rev.D 89, 029903 (2014)]"
}

@article{Kuhnel:2015vtw,
    author = {K{\"u}hnel, Florian and Rampf, Cornelius and Sandstad, Marit},
    title = "{Effects of Critical Collapse on Primordial Black-Hole Mass Spectra}",
    eprint = "1512.00488",
    archivePrefix = "arXiv",
    primaryClass = "astro-ph.CO",
    reportNumber = "NORDITA-2015-131",
    doi = "10.1140/epjc/s10052-016-3945-8",
    journal = "Eur. Phys. J. C",
    volume = "76",
    number = "2",
    pages = "93",
    year = "2016"
}

@article{Carr:2020xqk,
    author = "Carr, Bernard and Kuhnel, Florian",
    title = "{Primordial Black Holes as Dark Matter: Recent Developments}",
    eprint = "2006.02838",
    archivePrefix = "arXiv",
    primaryClass = "astro-ph.CO",
    doi = "10.1146/annurev-nucl-050520-125911",
    journal = "Ann. Rev. Nucl. Part. Sci.",
    volume = "70",
    pages = "355--394",
    year = "2020"
}

@article{Magana:2022cwq,
    author = "Maga{\~n}a, J. and San Mart{\'\i}n, M. and Sureda, J. and Rubio, M. and Araya, I. and Padilla, N.",
    title = "{Extended primordial black hole mass functions with a spike}",
    eprint = "2207.13689",
    archivePrefix = "arXiv",
    primaryClass = "astro-ph.CO",
    doi = "10.1093/mnras/stad261",
    journal = "Mon. Not. Roy. Astron. Soc.",
    volume = "520",
    number = "3",
    pages = "4276--4288",
    year = "2023"
}

@article{Kuhnel:2017pwq,
    author = {K{\"u}hnel, Florian and Freese, Katherine},
    title = "{Constraints on Primordial Black Holes with Extended Mass Functions}",
    eprint = "1701.07223",
    archivePrefix = "arXiv",
    primaryClass = "astro-ph.CO",
    doi = "10.1103/PhysRevD.95.083508",
    journal = "Phys. Rev. D",
    volume = "95",
    number = "8",
    pages = "083508",
    year = "2017"
}

@article{Green:2004wb,
    author = "Green, Anne M. and Liddle, Andrew R. and Malik, Karim A. and Sasaki, Misao",
    title = "{A New calculation of the mass fraction of primordial black holes}",
    eprint = "astro-ph/0403181",
    archivePrefix = "arXiv",
    doi = "10.1103/PhysRevD.70.041502",
    journal = "Phys. Rev. D",
    volume = "70",
    pages = "041502",
    year = "2004"
}

@article{Byrnes:2012yx,
    author = "Byrnes, Christian T. and Copeland, Edmund J. and Green, Anne M.",
    title = "{Primordial black holes as a tool for constraining non-Gaussianity}",
    eprint = "1206.4188",
    archivePrefix = "arXiv",
    primaryClass = "astro-ph.CO",
    reportNumber = "CERN-PH-TH-2012-167",
    doi = "10.1103/PhysRevD.86.043512",
    journal = "Phys. Rev. D",
    volume = "86",
    pages = "043512",
    year = "2012"
}

@article{Young:2013oia,
    author = "Young, Sam and Byrnes, Christian T.",
    title = "{Primordial black holes in non-Gaussian regimes}",
    eprint = "1307.4995",
    archivePrefix = "arXiv",
    primaryClass = "astro-ph.CO",
    doi = "10.1088/1475-7516/2013/08/052",
    journal = "JCAP",
    volume = "08",
    pages = "052",
    year = "2013"
}

@article{Atal:2018neu,
    author = "Atal, Vicente and Germani, Cristiano",
    title = "{The role of non-gaussianities in Primordial Black Hole formation}",
    eprint = "1811.07857",
    archivePrefix = "arXiv",
    primaryClass = "astro-ph.CO",
    reportNumber = "ICCUB-18-022",
    doi = "10.1016/j.dark.2019.100275",
    journal = "Phys. Dark Univ.",
    volume = "24",
    pages = "100275",
    year = "2019"
}

@article{Dolgov:1992pu,
    author = "Dolgov, Alexandre and Silk, Joseph",
    title = "{Baryon isocurvature fluctuations at small scales and baryonic dark matter}",
    reportNumber = "CFPA-TH-92-04",
    doi = "10.1103/PhysRevD.47.4244",
    journal = "Phys. Rev. D",
    volume = "47",
    pages = "4244--4255",
    year = "1993"
}

@article{Klein:1926tv,
    author = "Klein, Oskar",
    editor = "Taylor, J. C.",
    title = "{Quantum Theory and Five-Dimensional Theory of Relativity. (In German and English)}",
    doi = "10.1007/BF01397481",
    journal = "Z. Phys.",
    volume = "37",
    pages = "895--906",
    year = "1926"
}

@article{Arkani-Hamed:1998jmv,
    author = "Arkani-Hamed, Nima and Dimopoulos, Savas and Dvali, G. R.",
    title = "{The Hierarchy problem and new dimensions at a millimeter}",
    eprint = "hep-ph/9803315",
    archivePrefix = "arXiv",
    reportNumber = "SLAC-PUB-7769, SU-ITP-98-13",
    doi = "10.1016/S0370-2693(98)00466-3",
    journal = "Phys. Lett. B",
    volume = "429",
    pages = "263--272",
    year = "1998"
}

@article{Antoniadis:1998ig,
    author = "Antoniadis, Ignatios and Arkani-Hamed, Nima and Dimopoulos, Savas and Dvali, G. R.",
    title = "{New dimensions at a millimeter to a Fermi and superstrings at a TeV}",
    eprint = "hep-ph/9804398",
    archivePrefix = "arXiv",
    reportNumber = "SLAC-PUB-7801, SU-ITP-98-28, CPTH-S608-0498, IC-98-39",
    doi = "10.1016/S0370-2693(98)00860-0",
    journal = "Phys. Lett. B",
    volume = "436",
    pages = "257--263",
    year = "1998"
}

@article{Kaluza:1921tu,
    author = "Kaluza, Th.",
    title = {{Zum Unit{\"a}tsproblem der Physik}},
    eprint = "1803.08616",
    archivePrefix = "arXiv",
    primaryClass = "physics.hist-ph",
    reportNumber = "HUPD-8401",
    doi = "10.1142/S0218271818700017",
    journal = "Sitzungsber. Preuss. Akad. Wiss. Berlin (Math. Phys. )",
    volume = "1921",
    pages = "966--972",
    year = "1921"
}

@article{Overduin:1997sri,
    author = "Overduin, J. M. and Wesson, P. S.",
    title = "{Kaluza-Klein gravity}",
    eprint = "gr-qc/9805018",
    archivePrefix = "arXiv",
    doi = "10.1016/S0370-1573(96)00046-4",
    journal = "Phys. Rept.",
    volume = "283",
    pages = "303--380",
    year = "1997"
}

@article{Ananda:2006af,
    author = "Ananda, Kishore N. and Clarkson, Chris and Wands, David",
    title = "{The Cosmological gravitational wave background from primordial density perturbations}",
    eprint = "gr-qc/0612013",
    archivePrefix = "arXiv",
    doi = "10.1103/PhysRevD.75.123518",
    journal = "Phys. Rev. D",
    volume = "75",
    pages = "123518",
    year = "2007"
}

@article{Baumann:2007zm,
    author = "Baumann, Daniel and Steinhardt, Paul J. and Takahashi, Keitaro and Ichiki, Kiyotomo",
    title = "{Gravitational Wave Spectrum Induced by Primordial Scalar Perturbations}",
    eprint = "hep-th/0703290",
    archivePrefix = "arXiv",
    doi = "10.1103/PhysRevD.76.084019",
    journal = "Phys. Rev. D",
    volume = "76",
    pages = "084019",
    year = "2007"
}

@article{Kohri:2024qpd,
    author = "Kohri, Kazunori and Terada, Takahiro and Yanagida, Tsutomu T.",
    title = "{Induced gravitational waves probing primordial black hole dark matter with the memory burden effect}",
    eprint = "2409.06365",
    archivePrefix = "arXiv",
    primaryClass = "astro-ph.CO",
    reportNumber = "KEK-TH-2654, KEK-Cosmo-0358",
    doi = "10.1103/PhysRevD.111.063543",
    journal = "Phys. Rev. D",
    volume = "111",
    number = "6",
    pages = "063543",
    year = "2025"
}

@book{Appelquist:1987nr,
    editor = "Appelquist, T. and Chodos, A. and Freund, P. G. O.",
    title = "{MODERN KALUZA-KLEIN THEORIES}",
    year = "1987"
}

@book{Mukhanov:2005sc,
    author = "Mukhanov, V.",
    title = "{Physical Foundations of Cosmology}",
    doi = "10.1017/CBO9780511790553",
    isbn = "978-0-521-56398-7",
    publisher = "Cambridge University Press",
    address = "Oxford",
    year = "2005"
}

@book{Maggiore:2007ulw,
    author = "Maggiore, Michele",
    title = "{Gravitational Waves. Vol. 1: Theory and Experiments}",
    doi = "10.1093/acprof:oso/9780198570745.001.0001",
    isbn = "978-0-19-171766-6, 978-0-19-852074-0",
    publisher = "Oxford University Press",
    year = "2007"
}

@article{Inomata:2018epa,
    author = "Inomata, Keisuke and Nakama, Tomohiro",
    title = "{Gravitational waves induced by scalar perturbations as probes of the small-scale primordial spectrum}",
    eprint = "1812.00674",
    archivePrefix = "arXiv",
    primaryClass = "astro-ph.CO",
    reportNumber = "IPMU 18-0200",
    doi = "10.1103/PhysRevD.99.043511",
    journal = "Phys. Rev. D",
    volume = "99",
    number = "4",
    pages = "043511",
    year = "2019"
}

@article{Chapline:1975ojl,
    author = "Chapline, George F.",
    title = "{Cosmological effects of primordial black holes}",
    doi = "10.1038/253251a0",
    journal = "Nature",
    volume = "253",
    number = "5489",
    pages = "251--252",
    year = "1975"
}

@article{Carr:2023tpt,
    author = "Carr, Bernard and Clesse, Sebastien and Garcia-Bellido, Juan and Hawkins, Michael and Kuhnel, Florian",
    title = "{Observational evidence for primordial black holes: A positivist perspective}",
    eprint = "2306.03903",
    archivePrefix = "arXiv",
    primaryClass = "astro-ph.CO",
    doi = "10.1016/j.physrep.2023.11.005",
    journal = "Phys. Rept.",
    volume = "1054",
    pages = "1--68",
    year = "2024"
}

@article{Horowitz:2016lib,
    author = "Horowitz, Benjamin",
    title = "{Revisiting Primordial Black Holes Constraints from Ionization History}",
    eprint = "1612.07264",
    archivePrefix = "arXiv",
    primaryClass = "astro-ph.CO",
    month = "12",
    year = "2016"
}

@article{Ricotti:2007au,
    author = "Ricotti, Massimo and Ostriker, Jeremiah P. and Mack, Katherine J.",
    title = "{Effect of Primordial Black Holes on the Cosmic Microwave Background and Cosmological Parameter Estimates}",
    eprint = "0709.0524",
    archivePrefix = "arXiv",
    primaryClass = "astro-ph",
    doi = "10.1086/587831",
    journal = "Astrophys. J.",
    volume = "680",
    pages = "829",
    year = "2008"
}

@article{MACHO:2000qbb,
    author = "Alcock, C. and others",
    collaboration = "MACHO",
    title = "{The MACHO project: Microlensing results from 5.7 years of LMC observations}",
    eprint = "astro-ph/0001272",
    archivePrefix = "arXiv",
    doi = "10.1086/309512",
    journal = "Astrophys. J.",
    volume = "542",
    pages = "281--307",
    year = "2000"
}

@article{Niikura:2017zjd,
    author = "Niikura, Hiroko and others",
    title = "{Microlensing constraints on primordial black holes with Subaru/HSC Andromeda observations}",
    eprint = "1701.02151",
    archivePrefix = "arXiv",
    primaryClass = "astro-ph.CO",
    doi = "10.1038/s41550-019-0723-1",
    journal = "Nature Astron.",
    volume = "3",
    number = "6",
    pages = "524--534",
    year = "2019"
}

@article{Sasaki:2016jop,
    author = "Sasaki, Misao and Suyama, Teruaki and Tanaka, Takahiro and Yokoyama, Shuichiro",
    title = "{Primordial Black Hole Scenario for the Gravitational-Wave Event GW150914}",
    eprint = "1603.08338",
    archivePrefix = "arXiv",
    primaryClass = "astro-ph.CO",
    reportNumber = "RESCEU-17-16, RUP-16-7, YITP-16-43",
    doi = "10.1103/PhysRevLett.117.061101",
    journal = "Phys. Rev. Lett.",
    volume = "117",
    number = "6",
    pages = "061101",
    year = "2016",
    note = "[Erratum: Phys.Rev.Lett. 121, 059901 (2018)]"
}

@article{Lust:2019zwm,
    author = {L{\"u}st, Dieter and Palti, Eran and Vafa, Cumrun},
    title = "{AdS and the Swampland}",
    eprint = "1906.05225",
    archivePrefix = "arXiv",
    primaryClass = "hep-th",
    doi = "10.1016/j.physletb.2019.134867",
    journal = "Phys. Lett. B",
    volume = "797",
    pages = "134867",
    year = "2019"
}

@article{Kapner:2006si,
    author = "Kapner, D. J. and Cook, T. S. and Adelberger, E. G. and Gundlach, J. H. and Heckel, Blayne R. and Hoyle, C. D. and Swanson, H. E.",
    title = "{Tests of the gravitational inverse-square law below the dark-energy length scale}",
    eprint = "hep-ph/0611184",
    archivePrefix = "arXiv",
    doi = "10.1103/PhysRevLett.98.021101",
    journal = "Phys. Rev. Lett.",
    volume = "98",
    pages = "021101",
    year = "2007"
}

@article{Friedlander:2022ttk,
    author = "Friedlander, Avi and Mack, Katherine J. and Schon, Sarah and Song, Ningqiang and Vincent, Aaron C.",
    title = "{Primordial black hole dark matter in the context of extra dimensions}",
    eprint = "2201.11761",
    archivePrefix = "arXiv",
    primaryClass = "hep-ph",
    doi = "10.1103/PhysRevD.105.103508",
    journal = "Phys. Rev. D",
    volume = "105",
    number = "10",
    pages = "103508",
    year = "2022"
}

@article{Dvali:2024hsb,
    author = "Dvali, Gia and Valbuena-Berm{\'u}dez, Juan Sebasti{\'a}n and Zantedeschi, Michael",
    title = "{Memory burden effect in black holes and solitons: Implications for PBH}",
    eprint = "2405.13117",
    archivePrefix = "arXiv",
    primaryClass = "hep-th",
    doi = "10.1103/PhysRevD.110.056029",
    journal = "Phys. Rev. D",
    volume = "110",
    number = "5",
    pages = "056029",
    year = "2024"
}

@article{Keith:2020jww,
    author = "Keith, Celeste and Hooper, Dan and Blinov, Nikita and McDermott, Samuel D.",
    title = "{Constraints on Primordial Black Holes From Big Bang Nucleosynthesis Revisited}",
    eprint = "2006.03608",
    archivePrefix = "arXiv",
    primaryClass = "astro-ph.CO",
    reportNumber = "FERMILAB-PUB-20-224-A",
    doi = "10.1103/PhysRevD.102.103512",
    journal = "Phys. Rev. D",
    volume = "102",
    number = "10",
    pages = "103512",
    year = "2020"
}

@article{Du:2020rlx,
    author = "Du, Yuchen and Tahura, Shammi and Vaman, Diana and Yagi, Kent",
    title = "{Probing Compactified Extra Dimensions with Gravitational Waves}",
    eprint = "2004.03051",
    archivePrefix = "arXiv",
    primaryClass = "gr-qc",
    doi = "10.1103/PhysRevD.103.044031",
    journal = "Phys. Rev. D",
    volume = "103",
    number = "4",
    pages = "044031",
    year = "2021"
}

@article{Antoniadis:2023sya,
    author = "Antoniadis, Ignatios and Cunat, Jules and Guillen, Anthony",
    title = "{Cosmological perturbations from five-dimensional inflation}",
    eprint = "2311.17680",
    archivePrefix = "arXiv",
    primaryClass = "hep-ph",
    doi = "10.1007/JHEP05(2024)290",
    journal = "JHEP",
    volume = "05",
    pages = "290",
    year = "2024"
}

@article{Korwar:2023kpy,
    author = "Korwar, Mrunal and Profumo, Stefano",
    title = "{Updated constraints on primordial black hole evaporation}",
    eprint = "2302.04408",
    archivePrefix = "arXiv",
    primaryClass = "hep-ph",
    doi = "10.1088/1475-7516/2023/05/054",
    journal = "JCAP",
    volume = "05",
    pages = "054",
    year = "2023"
}

@article{Tashiro:2008sf,
    author = "Tashiro, Hiroyuki and Sugiyama, Naoshi",
    title = "{Constraints on Primordial Black Holes by Distortions of Cosmic Microwave Background}",
    eprint = "0801.3172",
    archivePrefix = "arXiv",
    primaryClass = "astro-ph",
    doi = "10.1103/PhysRevD.78.023004",
    journal = "Phys. Rev. D",
    volume = "78",
    pages = "023004",
    year = "2008"
}

@article{Bird:2010mp,
    author = "Bird, Simeon and Peiris, Hiranya V. and Viel, Matteo and Verde, Licia",
    title = "{Minimally Parametric Power Spectrum Reconstruction from the Lyman-alpha Forest}",
    eprint = "1010.1519",
    archivePrefix = "arXiv",
    primaryClass = "astro-ph.CO",
    doi = "10.1111/j.1365-2966.2011.18245.x",
    journal = "Mon. Not. Roy. Astron. Soc.",
    volume = "413",
    pages = "1717--1728",
    year = "2011"
}

@article{Fixsen:1996nj,
    author = "Fixsen, D. J. and Cheng, E. S. and Gales, J. M. and Mather, John C. and Shafer, R. A. and Wright, E. L.",
    title = "{The Cosmic Microwave Background spectrum from the full COBE FIRAS data set}",
    eprint = "astro-ph/9605054",
    archivePrefix = "arXiv",
    doi = "10.1086/178173",
    journal = "Astrophys. J.",
    volume = "473",
    pages = "576",
    year = "1996"
}

@article{Lee:2020wfn,
    author = "Lee, Vincent S. H. and Mitridate, Andrea and Trickle, Tanner and Zurek, Kathryn M.",
    title = "{Probing Small-Scale Power Spectra with Pulsar Timing Arrays}",
    eprint = "2012.09857",
    archivePrefix = "arXiv",
    primaryClass = "astro-ph.CO",
    doi = "10.1007/JHEP06(2021)028",
    journal = "JHEP",
    volume = "06",
    pages = "028",
    year = "2021"
}

@article{NANOGrav:2023gor,
    author = "Agazie, Gabriella and others",
    collaboration = "NANOGrav",
    title = "{The NANOGrav 15 yr Data Set: Evidence for a Gravitational-wave Background}",
    eprint = "2306.16213",
    archivePrefix = "arXiv",
    primaryClass = "astro-ph.HE",
    doi = "10.3847/2041-8213/acdac6",
    journal = "Astrophys. J. Lett.",
    volume = "951",
    number = "1",
    pages = "L8",
    year = "2023"
}

@article{NANOGrav:2023hvm,
    author = "Afzal, Adeela and others",
    collaboration = "NANOGrav",
    title = "{The NANOGrav 15 yr Data Set: Search for Signals from New Physics}",
    eprint = "2306.16219",
    archivePrefix = "arXiv",
    primaryClass = "astro-ph.HE",
    reportNumber = "FERMILAB-PUB-23-589-T",
    doi = "10.3847/2041-8213/acdc91",
    journal = "Astrophys. J. Lett.",
    volume = "951",
    number = "1",
    pages = "L11",
    year = "2023",
    note = "[Erratum: Astrophys.J.Lett. 971, L27 (2024), Erratum: Astrophys.J. 971, L27 (2024)]"
}

@article{Chluba:2019kpb,
    author = "Chluba, J. and others",
    title = "{Spectral Distortions of the CMB as a Probe of Inflation, Recombination, Structure Formation and Particle Physics}: {Astro2020 Science White Paper}",
    eprint = "1903.04218",
    archivePrefix = "arXiv",
    primaryClass = "astro-ph.CO",
    journal = "Bull. Am. Astron. Soc.",
    volume = "51",
    number = "3",
    pages = "184",
    year = "2019"
}

@article{A_Kogut_2011,
   title={The Primordial Inflation Explorer (PIXIE): a nulling polarimeter for cosmic microwave background observations},
   volume={2011},
   ISSN={1475-7516},
   url={http://dx.doi.org/10.1088/1475-7516/2011/07/025},
   DOI={10.1088/1475-7516/2011/07/025},
   number={07},
   journal={Journal of Cosmology and Astroparticle Physics},
   publisher={IOP Publishing},
   author={A. Kogut and D.J. Fixsen and D.T. Chuss and J. Dotson and E. Dwek and M. Halpern and G.F. Hinshaw and S.M. Meyer and S.H. Moseley and M.D. Seiffert and D.N. Spergel and E.J. Wollack},
   year={2011},
   month=jul, pages={025–025} }

@article{LISA:2017pwj,
    author = "Amaro-Seoane, Pau and others",
    collaboration = "LISA",
    title = "{Laser Interferometer Space Antenna}",
    eprint = "1702.00786",
    archivePrefix = "arXiv",
    primaryClass = "astro-ph.IM",
    month = "2",
    year = "2017"
}

@article{Crowder:2005nr,
    author = "Crowder, Jeff and Cornish, Neil J.",
    title = "{Beyond LISA: Exploring future gravitational wave missions}",
    eprint = "gr-qc/0506015",
    archivePrefix = "arXiv",
    doi = "10.1103/PhysRevD.72.083005",
    journal = "Phys. Rev. D",
    volume = "72",
    pages = "083005",
    year = "2005"
}

@article{Seto:2001qf,
    author = "Seto, Naoki and Kawamura, Seiji and Nakamura, Takashi",
    title = "{Possibility of direct measurement of the acceleration of the universe using 0.1-Hz band laser interferometer gravitational wave antenna in space}",
    eprint = "astro-ph/0108011",
    archivePrefix = "arXiv",
    doi = "10.1103/PhysRevLett.87.221103",
    journal = "Phys. Rev. Lett.",
    volume = "87",
    pages = "221103",
    year = "2001"
}

@article{Punturo:2010zz,
    author = "Punturo, M. and others",
    editor = "Ricci, Fulvio",
    title = "{The Einstein Telescope: A third-generation gravitational wave observatory}",
    doi = "10.1088/0264-9381/27/19/194002",
    journal = "Class. Quant. Grav.",
    volume = "27",
    pages = "194002",
    year = "2010"
}

@article{LIGOScientific:2016wof,
    author = "Abbott, Benjamin P and others",
    collaboration = "LIGO Scientific",
    title = "{Exploring the Sensitivity of Next Generation Gravitational Wave Detectors}",
    eprint = "1607.08697",
    archivePrefix = "arXiv",
    primaryClass = "astro-ph.IM",
    reportNumber = "LIGO-P1600143",
    doi = "10.1088/1361-6382/aa51f4",
    journal = "Class. Quant. Grav.",
    volume = "34",
    number = "4",
    pages = "044001",
    year = "2017"
}

@article{AEDGE:2019nxb,
    author = "El-Neaj, Yousef Abou and others",
    collaboration = "AEDGE",
    title = "{AEDGE: Atomic Experiment for Dark Matter and Gravity Exploration in Space}",
    eprint = "1908.00802",
    archivePrefix = "arXiv",
    primaryClass = "gr-qc",
    reportNumber = "KCL-PH-TH/2019-65, CERN-TH-2019-126",
    doi = "10.1140/epjqt/s40507-020-0080-0",
    journal = "EPJ Quant. Technol.",
    volume = "7",
    pages = "6",
    year = "2020"
}

@article{Janssen:2014dka,
    author = "Janssen, Gemma and others",
    editor = "Bourke, Tyler L. and others",
    title = "{Gravitational wave astronomy with the SKA}",
    eprint = "1501.00127",
    archivePrefix = "arXiv",
    primaryClass = "astro-ph.IM",
    doi = "10.22323/1.215.0037",
    journal = "PoS",
    volume = "AASKA14",
    pages = "037",
    year = "2015"
}

@article{Garcia-Bellido:2021zgu,
    author = "Garcia-Bellido, Juan and Murayama, Hitoshi and White, Graham",
    title = "{Exploring the early Universe with Gaia and Theia}",
    eprint = "2104.04778",
    archivePrefix = "arXiv",
    primaryClass = "hep-ph",
    reportNumber = "IFT-UAM/CSIC-2021-038",
    doi = "10.1088/1475-7516/2021/12/023",
    journal = "JCAP",
    volume = "12",
    number = "12",
    pages = "023",
    year = "2021"
}

@article{Acharya:2020jbv,
    author = "Acharya, Sandeep Kumar and Khatri, Rishi",
    title = "{CMB and BBN constraints on evaporating primordial black holes revisited}",
    eprint = "2002.00898",
    archivePrefix = "arXiv",
    primaryClass = "astro-ph.CO",
    doi = "10.1088/1475-7516/2020/06/018",
    journal = "JCAP",
    volume = "06",
    pages = "018",
    year = "2020"
}

@article{LIGOScientific:2025slb,
    author = "Abac, A. G. and others",
    collaboration = "LIGO Scientific, VIRGO, KAGRA",
    title = "{GWTC-4.0: Updating the Gravitational-Wave Transient Catalog with Observations from the First Part of the Fourth LIGO-Virgo-KAGRA Observing Run}",
    eprint = "2508.18082",
    archivePrefix = "arXiv",
    primaryClass = "gr-qc",
    reportNumber = "LIGO-P2400386",
    month = "8",
    year = "2025"
}

@article{Anchordoqui:2025nmb,
    author = "Anchordoqui, Luis A. and Antoniadis, Ignatios and Lust, Dieter",
    title = "{Two Micron-Size Dark Dimensions}",
    eprint = "2501.11690",
    archivePrefix = "arXiv",
    primaryClass = "hep-th",
    reportNumber = "MPP-2025-5, LMU-ASC 02/25",
    doi = "10.1002/prop.70015",
    journal = "Fortsch. Phys.",
    volume = "73",
    number = "8",
    pages = "e70015",
    year = "2025"
}

@article{Leontaris:2026kvu,
    author = "Leontaris, George K. and Prampromis, George",
    title = "{Micron-sized Extra Dimensions and Primordial Black Holes: Charged, Rotating, and Memory Burdened}",
    eprint = "2605.00252",
    archivePrefix = "arXiv",
    primaryClass = "hep-ph",
    month = "4",
    year = "2026"
}

@article{Vitale:2026ric,
    author = "Vitale, Giuseppe Filiberto and Lambiase, Gaetano and Poddar, Tanmay Kumar and Visinelli, Luca",
    title = "{Microscopic primordial black holes as macroscopic dark matter from large extra dimensions}",
    eprint = "2604.14871",
    archivePrefix = "arXiv",
    primaryClass = "astro-ph.CO",
    month = "4",
    year = "2026"
}

@article{Hardy:2025ajb,
    author = "Hardy, Edward and Sokolov, Anton and Stubbs, Henry",
    title = "{Stellar cooling limits on KK gravitons and dark dimensions}",
    eprint = "2510.18975",
    archivePrefix = "arXiv",
    primaryClass = "hep-ph",
    doi = "10.1007/JHEP03(2026)029",
    journal = "JHEP",
    volume = "03",
    pages = "029",
    year = "2026"
}

\end{document}